\documentclass[prd,aps,a4paper,eqsecnum,mssymb,nofootinbib,floatfix,twocolumn]{revtex4}  

\newif\ifusesec
\usesectrue    
\usepackage{graphicx} 
\usepackage{amsmath,amsfonts,amssymb}

\newcommand{\beq}{\begin{equation}}
\newcommand{\eeq}{\end{equation}}
\newcommand{\be}{\begin{equation}}
\newcommand{\ee}{\end{equation}}
\newcommand{\bea}{\begin{eqnarray}}
\newcommand{\eea}{\end{eqnarray}}
\newcommand{\eq}{\eqref}
\newcommand{\g}{{\gamma}}
\newcommand{\Ef}{{E}_{\rm eff}}
\newcommand{\hEf}{\widehat{E}_{\rm eff}}
\newcommand{\pinf}{p_{\infty}}
\newcommand{\D}{\partial}

\begin{document}

\title{Fourth Post-Minkowskian Local-in-Time Conservative  Dynamics of Binary Systems}

\author{Donato Bini$^{1,2}$, Thibault Damour$^3$}
  \affiliation{
$^1$Istituto per le Applicazioni del Calcolo ``M. Picone,'' CNR, I-00185 Rome, Italy\\
$^2$INFN, Sezione di Roma Tre, I-00146 Rome, Italy\\
$^3$Institut des Hautes Etudes Scientifiques, 91440 Bures-sur-Yvette, France
}
\date{\today}

\begin{abstract}
We compute the purely local-in-time (scale-free and logarithm-free) part of the conservative dynamics 
of gravitationally interacting
two-body systems at the fourth post-Minkowskian order, and at the thirtiest order in velocity. The gauge-invariant content
of this fourth post-Minkowskian local dynamics is given in two ways: (i) its contribution to the on-shell action (for both
hyperboliclike and ellipticlike motions); and (ii) its contribution to the Effective One Body Hamiltonian (in energy gauge).
Our computation capitalizes on the Tutti Frutti approach [Phys. Rev. Lett. \textbf{123}, no.23, 231104 (2019)],
and on recent post-Minkowskian advances [Phys. Rev. Lett. \textbf{128}, no.16, 161103 (2022)], [Phys. Rev. Lett. \textbf{128}, no.16, 161104 (2022)], and [Phys. Rev. Lett. \textbf{132}, no.22, 221401 (2024)].
\end{abstract}

\maketitle

\section{Introduction}

A novel approach to the conservative dynamics of gravitationally interacting binary systems (dubbed {\it Tutti Frutti} (TF)) has
been introduced in Ref. \cite{Bini:2019nra} and developed up to the sixth post-Newtonian (6PN) order in Refs. \cite{Bini:2020wpo,Bini:2020nsb}.
The Tutti Frutti (TF) approach combines information from different approximation methods: post-Newtonian (PN), post-Minkowskian (PM), Multipolar-post-Minkowskian (MPM), Self-Force, Effective-One-Body (EOB), and Delaunay averaging.  It also makes a crucial use of the
restricted dependence of the PM-expansion coefficients of various physical quantities on the symmetric mass ratio 
$\nu \equiv \frac{m_1 m_2}{(m_1+m_2)^2}$. In particular, the conservative scattering angle (considered as
a function of the impact parameter) was shown \cite{Damour:2019lcq} to depend essentially linearly on $\nu$ at the third and fourth PM orders.
More precisely, the PM-expansion of the  conservative scattering angle has the structure\footnote{Though we generally set $c=1$, we sometimes indicate the dependence on $c$ when it it clarifying.}
\beq 
\label{chiPM}
\frac{\chi^{\rm cons}(\g, b)}{2 h(\g,\nu)} = \sum_{n\geq 1} \tilde \chi_n(\g,\nu) \left( \frac{G M}{c^2  b \,\pinf}\right)^n\,,
\eeq
where $\tilde \chi_n(\g,\nu)$ is a polynomial in $\nu$ of order $\bar{n} \equiv \left[ \frac{n-1}{2} \right]$, 
which starts with the test-particle (or Schwarzschild) result, say 
\beq 
\label{tildechi}
 \tilde \chi_n(\g,\nu)= \chi_n^S+  {\chi_n}^{\nu^1} + \cdots +  {\chi_n}^{\nu^{\bar{n}}} \,.
\eeq
Everywhere in this paper the notation ${Q}^{\nu^n}$ means the term proportional to $\nu^n$ in the quantity $Q$.

In Eq. \eqref{tildechi}
$\g \equiv - (p_1 \cdot p_2)/(m_1 m_2 c^2)$ is the asymptotic relative Lorentz factor (we use the mostly positive signature for the spacetime metric), $b$ the impact parameter,
$M \equiv m_1+m_2$, $\pinf \equiv \sqrt{\g^2-1}$, and
\be \label{defh}
h(\g,\nu) \equiv \sqrt{1+ 2 \nu(\g-1)} = \frac{E}{M c^2}\,.
\ee 
In the last equation $E$ denotes the total center-of-mass (cm) energy of the system.
The Lorentz factor $\g$ can be parametrized either by $\pinf \equiv \sqrt{\g^2-1}$ or by $v \equiv \sqrt{1-\frac1{\g^2}}$ so that
\beq
\g = \sqrt{1+ \pinf^2} = \frac1{\sqrt{1-v^2}}.
\eeq

When $n= 3$ or $4$, one has $\bar{n}=1$ so that Eq. \eq{tildechi} says that $ \tilde \chi_n(\g,\nu)$ is linear in $\nu$.
This fact allows one, in principle, to derive the exact 4PM ($G^4$) conservative dynamics from {\it first-order} SF results.
This is illustrated in Fig. 1 of Ref. \cite{Bini:2020nsb}, where  the (third and) fourth column(s), 
labelling the velocity expansion of the ($(GM/r)^3$ and) $(GM/r)^4$ contributions to the local-in-time dynamics,
contain only two black dots (corresponding to $\nu^0$ and $\nu^1$).

In our previous works \cite{Bini:2019nra,Bini:2020wpo,Bini:2020nsb} we were interested in the PN expansion (rather than the PM one)
of the binary dynamics, which meant
considering contributions $\propto u^k p^{2 \ell}$ having a given value of the PN order $n^{\rm PN} \equiv k+\ell -1$.
[Each PN order corresponds to an hyphenated diagonal line in Fig. 1 of Ref. \cite{Bini:2020nsb}.
E.g., the first post-Newtonian (1PN)
 order, $n^{\rm PN}=1$, corresponds to  $p^4$, $ u p^2$ and $u^2$ terms.]
Our 6PN-order results \cite{Bini:2020nsb}  comprised terms from $p^{14}$ to $u^7 p^0$. In other words, they included
terms up to the 7PM order ($G^7$), the latter being computed to zeroth order in velocity. By contrast, the 4PM ($u^4 \sim G^4$) terms in the dynamics were computed to sixth order in velocity ($p^6$).

There has recently been an impressive progress on the PM precision-frontier using either 
quantum scattering amplitudes (notably \cite{Bern:2021dqo,Bern:2021yeh,Damgaard:2023ttc,Bern:2024adl}) or worldline approaches
(notably \cite{Dlapa:2021vgp,Jakobsen:2023ndj,Dlapa:2024cje,Driesse:2024xad}). The usefulness of  PM results
when describing the classical dynamics of binary systems has been pointed out in 
Refs. \cite{Damour:2016gwp,Damour:2017zjx,Cheung:2018wkq},
and explored in many subsequent works (see, e.g., \cite{Bini:2018ywr,Antonelli:2019ytb,
Cristofoli:2019neg,Kalin:2019rwq,Khalil:2022ylj,Damour:2022ybd}).

We capitalize here on the  PM results on
the 4PM conservative dynamics \cite{Bern:2021yeh,Dlapa:2021vgp},
and on a recent computation of the nonlocal-in-time tail part of the $G^4$  action \cite{Dlapa:2024cje},
and combine them with some results of the TF approach.
Specifically, the aim of the present work is: (i) to write down explicitly the integral ($W_2^{G^4}$) needed to 
obtain the purely local-in-time (scale-free and logarithm-free) conservative local-in-time 4PM dynamics;
(ii) to compute the latter integral to a very high order in velocity ($p^{30}$);
and (iii) to combine our new result with the 4PM knowledge of the conservative scattering angle \cite{Bern:2021yeh,Dlapa:2021vgp},
and with the recently acquired  \cite{Dlapa:2024cje} 
knowledge of the other  integral ($W_1^{G^4}$) entering the nonlocal 4PM dynamics, to  compute two gauge-invariant functions defining the 4PM  {\it purely local-in-time} dynamics: the $O(G^4)$ radial action $I_3$ and the
 $O(G^4)$ local Effective One-Body Hamiltonian. By contrast with the corresponding results of Ref. \cite{Dlapa:2024cje}
 (which notably contain logarithms having their origin in non-local tail effects), our results are free of logarithms.
 The results presented here allow us to reach a much higher PN accuracy than the one (6PN) reached in  our previous works.
We leave to future work the task of completing our results by computing the exact 4PM value of $W_2^{G^4}$
instead of its $O(p^{30})$ expansion presented here.

\section{Reminder of the structure of the Tutti Frutti (TF) conservative dynamics}

Starting from 4PN and 4PM orders [i.e., from $O(G^4/c^8)$], the conservative dynamics of a two body system must be described
by a nonlocal-in-time action \cite{Blanchet:1987wq,Foffa:2011np,Damour:2014jta,Damour:2015isa,Galley:2015kus,Bernard:2017bvn,Foffa:2019yfl,Bini:2020hmy}.  The TF approach decomposes in a specific manner the two-body action in nonlocal and local
parts, namely
\beq \label{Stot}
S_{\rm tot} = S_{\rm nonloc, f}+ S_{\rm loc,f}\,.
\eeq
Here, the additional subscript $f$ refers to the use of a {\it flexibility} factor, $f(t)$, whose role is discussed below.

Let us successively recall the definitions of the two parts (nonlocal-in-time and local-in-time) of the action Eq. \eq{Stot}.

\subsection{Nonlocal-in-time part of the TF action}

The explicit expression of  $S_{\rm nonloc, f}$ depends on the considered PM approximation order. Indeed, the 
nonlocality in time is linked to tail effects, which are described (say in the MPM formalism)
by the backscattering of retarded (and advanced) waves on the $O(\frac{GE}{r})$ part of the gravitational field \cite{Blanchet:1987wq}.
This can be computed as a perturbative expansion in $GE$, the leading-order (LO) tail involving one factor $GE$,
the next-to-leading-order (NLO) tail involving a factor  $(GE)^2$, etc.  The LO tail-transported nonlocality is
of order $\frac{G^4}{c^8}(1+ \frac1{c^2}+ \cdots)$, i.e. at 4PM and starting at 4PN. The  NLO tail-transported nonlocality 
has been discussed in \cite{Damour:2015isa,Bini:2020wpo}. 

It starts at the order  $\frac{G^5}{c^{11}}$ (i.e., 5PM and 5.5PN). As our present work
limits itself to the 4PM, $O(G^4)$, approximation, we only consider linear-tail nonlocal interactions. These give rise
to a nonlocal action which is written, in the TF approach, in the form
\beq 
\label{Snonloc}
 S_{\rm nonloc, f}= - \int dt  H_{\rm nonloc, f}(t)\,,
\eeq
with a {\it flexed} nonlocal Hamiltonian given by
\beq 
\label{Hnonloc0}
 H_{\rm nonloc, f}(t) = - \frac{ G E}{c^5} \, {\rm Pf}_{2r^f_{12}/c}\int \frac{dt'}{|t-t'|}{\mathcal F}_{\rm cm}^{\rm split}(t,t')\,,
\eeq
where ${\rm Pf}_T$ denotes a (Hadamard) partie finie of the logarithmically divergent (near $t'=t$) integral over $t'$,
regularized with the time scale $T$, and 
where ${\mathcal F}_{\rm cm}^{\rm split}(t,t')$ denotes the ``time-split" version of the gravitational-wave  energy flux (in the
cm system). The function ${\mathcal F}_{\rm cm}^{\rm split}(t,t')$ is such that its
 diagonal value (i.e. when $t=t'$)  coincides with the
instantaneous gravitational-wave  energy flux:  ${\mathcal F}_{\rm cm}^{\rm split}(t,t)=  {\mathcal F}_{\rm cm}(t)$.

The structure Eq. \eq{Hnonloc0} of the nonlocal-in-time Hamiltonian has been derived by
a series of complementary approaches \cite{Blanchet:1987wq,Foffa:2011np,Damour:2014jta,Damour:2015isa,Galley:2015kus,Bini:2020hmy}.

In the MPM formalism, the integrand ${\mathcal F}_{\rm cm}^{\rm split}(t,t')$
is originally defined in terms of the (even-parity and odd-parity) source multipole moments,
$I_L(t)$, $J_L(t)$ (with $L=i_1 i_2 \cdots i_\ell$) of the system  by
\bea 
\label{Fsplit1}
{\mathcal F}_{\rm cm}^{\rm split}(t,t') &=& \sum_{\ell \geq 2}\left[ c^I_\ell I^{(\ell+1)}_L(t) I_L^{(\ell+1)}(t')\right.\nonumber\\
&+&\left. c^J_\ell J^{(\ell+1)}_L(t) J^{(\ell+1)}_L(t')\right]\,,
\eea
with the same numerical coefficients (starting with $ c^I_2=\frac15$) entering the multipole
expansion of the instantaneous gravitational-wave  energy flux (considered in the cm system), so that
\bea
{\mathcal F}_{\rm cm}(t)&=&{\mathcal F}_{\rm cm}^{\rm split}(t,t) \nonumber\\
&=& \sum_{\ell \geq 2}\left[ c^I_\ell I_L^{(\ell+1)}(t) I^{(\ell+1)}_L(t)\right.\nonumber\\
&+&\left. c^J_\ell J_L^{(\ell+1)}(t) J_L^{(\ell+1)}(t)\right]\,.
\eea
The above expressions have been found to give results in agreement with corresponding results derived from
self-force computations up to at least the order $G^7/c^{12}$ reached in Ref. \cite{Bini:2020hmy}.

The definition Eq. \eq{Fsplit1} is equivalent to (with $ d\Omega_{\rm cm}=  \sin \theta_{\rm cm} d \theta_{\rm cm} d \phi_{\rm cm}$ )
\begin{widetext}
\bea
\label{Fttp_Waveform}
{\mathcal F}_{\rm cm}(t,t')&=& \frac{1}{16 \pi G}\int d\Omega_{\rm cm}\left[\frac{df_c(u, \theta_{\rm cm}, \phi_{\rm cm})}{du}\right]^{u=t} \left[ \frac{d f_c^*(u', \theta_{\rm cm}, \phi_{\rm cm})}{du'}\right]^{u'=t'}\,,
\eea 
\end{widetext}
where $f_c(u, \theta_{\rm cm}, \phi_{\rm cm})= \lim_{r \to \infty} r  (h_+ +i h_\times)$ denotes the asymptotic waveform in the cm frame. Here, $u$ (before its identification with the time variable of the nonlocal action)
denotes the cm-frame retarded time $u  \simeq t - \frac{r}{c} -  \frac{2 G E}{c^5} \ln \left( \frac{r}{r_0} \right)$.

An important aspect of the TF definition of the nonlocal Hamiltonian, Eq. \eqref{Hnonloc0}, is the use of the specific
time scale $2r^f_{12}(t)/c$, where
\be \label{r12f}
r^f_{12}(t) = f(t) r^h_{12}(t)\,,
\ee
differs from the harmonic-coordinate  cm relative distance, $r^h_{12}(t)$, by a flexibility factor $f(t)$ satisfying
\beq 
\label{fnu}
f(t) = 1 + O(\nu)\,.
\eeq

The use of such a regularization scale has two effects: First, the fact that it is anchored on the distance
$ r^h_{12}$ between the two bodies has the property (already emphasized in Ref. \cite{Damour:2014jta}) of freeing the corresponding
local-in-time Hamiltonian from the presence of any logarithm.  Let us recall the origin of logarithms when dividing
the study of binary systems in two complementary regions: a near-zone region considered up to some 
infrared cut-off length scale 
$s \gg r_{12}$, and a complementary radiative-zone region  $|{\bf x}| > s$. Here, the length scale $s$ can be
physically taken as being intermediate between $r_{12}$ and  the wavelength  $\lambda \sim r_{12} c/v_{12}$ 
of the emitted radiation ($v_{12}$ being a characteristic velocity, formally assumed to be small).
The appearance of logarithms $\sim \ln (r_{12}/s)$ in the near-zone-only\footnote{Corresponding to
potential-graviton interactions.} part of the dynamics  is linked to infrared divergences associated with nonlocal tail effects.
 The inclusion of radiative-zone interactions adds logarithmic terms $\sim -\ln(\lambda/s)$. When adding the two contributions the dependence on the
intermediate scale $r_{12} \ll s \ll \lambda$ cancels out to leave (in the small velocity limit)
a logarithm of the velocity $\ln(\frac{c}{v_{12}})$. The TF formalism considers that both logarithms,  
$ \ln (r_{12}/s)$ and $ \ln(s/\lambda)$, most naturally belong
to the nonlocal part of the dynamics, and this is achieved by anchoring the partie-finie regularization scale 
entering $S_{\rm nonloc, f}$ on $r_{12}$.

The second effect of the choice of regularization scale, Eq. \eq{r12f},  is linked to the flexibility factor $f(t)$.
We recalled above that the (total) scattering angle (derived from the full, local+nonlocal, action) must satisfy
the simple $\nu$-dependence displayed in Eq. \eq{tildechi}. The TF formalism makes the natural choice of
defining the  local+nonlocal split of the action so that the corresponding partial scattering angles
$\chi^{\rm loc}$ and $\chi^{\rm nonloc}$, with
\be \label{chisplit}
\chi^{\rm tot} =\chi^{\rm loc} + \chi^{\rm nonloc} + O(G^8)\,,
\ee
separately satisfy the simple $\nu$-dependence displayed in Eq. \eq{tildechi}. The $ O(G^8)$ error term in the
above equation is due to the fact that we treat here the $O(G^4)$ nonlocal action as a linear perturbation of the
local action $S_{\rm loc, f}$. When working at the 6PN approximation, Refs. \cite{Bini:2020nsb,Bini:2020hmy} have explicitly shown the
possibility of defining a flexibility factor $f(t) = 1 + O(\nu)$ such that both $\chi^{\rm loc}$ and $\chi^{\rm nonloc}$
satisfy Eq. \eq{tildechi}. We shall show below that this is also possible when considering the $G^4$ dynamics
at all velocity orders.

\subsection{Local-in-time part of the TF action}

One of the advantages of the decomposition Eq. \eq{Stot} is that, contrary to the nonlocal part, $S_{\rm nonloc, f}$,
the local part of the action, $S_{\rm loc, f}$, can be described by a usual Hamiltonian  that can be 
uniformly applied to scattering states and bound states. The latter local Hamiltonian  admits a normal perturbative
description (which can be either of the PN type or of the PM type).

 The full gauge-invariant information contained in the local dynamics can be encoded in various ways,
 depending also whether one is interested in scattering states or bound states.
 
When considering scattering states, a first way is to give the rescaled PM-expansion coefficients $\tilde \chi^{\rm loc}$ (see Eq. \eqref{chiPM})
of the local-in-time contribution to the scattering angle. Equivalently, one can give the (regularized) hyperbolic radial action,
$I_r^{\rm hyp , loc}$, giving rise to the (full) local deflection angle via
its dependence on the total (cm) angular momentum $J$:
\be
\pi + \chi^{\rm loc} = - \frac{\D I_r^{\rm hyp , loc}}{\D J}= -  \frac{\D \widehat I_r^{\rm hyp , loc}}{\D j}\,.
\ee
Here, the second equation uses the dimensionless versions of $I_r^{\rm hyp}$ and $J$, namely
${\widehat I_r}^{\rm hyp , loc} \equiv \frac{I_r^{\rm hyp}}{G m_1 m_2}$ and $j \equiv \frac{  J}{Gm_1 m_2}$.

In view of Eq. \eq{chiPM}, and of the relation
\be
\frac1j= \frac{G E}{ b \pinf}= \frac{GM h}{b \pinf}\,,
\ee
the PM expansion of the radial action reads 
\bea
\label{Ir_hyp_loc}
{\widehat I_r}^{\rm hyp , loc} &=& \frac{I_r^{\rm hyp}}{G m_1 m_2}\nonumber\\
&=& - \pi j +  2 \frac{2 \g^2-1}{\sqrt{\g^2-1}} \ln \frac{j_0(\g,\nu)}{j}\nonumber\\
&+&\sum_{n\geq 1}  \frac{ {\widehat I_{r,n}}^{\rm hyp , loc} (\g,\nu) }{(h \,j)^n}\,,
\eea
where $j_0(\g,\nu)$ is an arbitrary function of $\g$ and $\nu$ linked to the regularization of the (Coulomb) logarithmic
divergence of the radial action, and where
\bea
{\widehat I}_{r,n}^{\rm hyp , loc} (\g,\nu) &=&   \frac{2}{n} \tilde \chi^{\rm loc}_{n+1} (\g,\nu)\,.
\eea
Note that ${\widehat I}_{r,n}^{\rm hyp , loc} (\g,\nu)$ is a polynomial in $\nu$ of order $\left[ \frac{n}{2}\right]$,
and that a term in the radial action $\sim \frac1{j^n}$ corresponds to the $O(G^{n+1})$ level.
At the 4PM level ${\widehat I}_{r,3}^{\rm hyp , loc}(\g,\nu) =   \frac{2}{3} \tilde \chi^{\rm loc}_{4} (\g,\nu)$
is linear in $\nu$, i.e.,
\be
{\widehat I}_{r,3}^{\rm hyp , loc}(\g,\nu) = \widehat{ I}_{r, 3, S}^{\rm hyp , loc}(\g)+ \widehat{ I}_{r, 3}^{\rm hyp , loc, \nu^1}(\g)\,,
\ee
where 
\be
\label{schw_Ir3}
 \widehat{ I}_{r, 3, S}^{\rm hyp , loc}(\g) = \frac{\pi }{64}(1155 \g^4 - 630 \g^2 +35) 
\ee
is the Schwarzschild (probe) contribution. The full gauge-invariant content of the local 4PM dynamics is therefore
contained in the linear-in-$\nu$, $G m_1m_2$-rescaled radial action $\widehat{ I}_{r, 3}^{\rm hyp , loc, \nu^1}(\g)$.
When considering the unrescaled radial action $G m_1m_2 \widehat{ I}_{r, 3}^{\rm hyp , loc, \nu^1}(\g)$ the
 gauge-invariant content of the latter contribution is proportional to $\nu^2$.

The split Eq. \eq{chisplit}  of the scattering angle in local and nonlocal pieces implies a corresponding split of
the radial action,
\beq 
\label{Irtot}
I_r^{\rm  hyp, tot} =I_r^{\rm hyp, loc} + I_r^{\rm hyp, nonloc} \,.
\eeq
In view of the results of  Ref. \cite{Bini:2020hmy}, the nonlocal part of the radial action can be written
in terms of the nonlocal part of the Hamiltonian (treated as a perturbation of the local part) as
\beq
 I_r^{\rm hyp, nonloc} = - \left[ \int dt H_{\rm nonloc} \right]^{\rm on-shell}\,,
 \eeq
 where the superscript ``on-shell" means that it is integrated along an hyperboliclike solution of the  dynamics
 defined by the local Hamiltonian. [This differs from Eq. \eq{Snonloc} which referred to an action principle, i.e.
 a functional to be varied.]
 
 Inserting Eq. \eq{Hnonloc0}, and introducing an arbitrary intermediate length scale $s$, then leads to
 \beq 
\label{IrW}
  I_r^{\rm hyp, nonloc} = - W_1(s) - W_2(s)- W_{\rm  f-h}\,,
 \eeq
 where
\bea \label{Ws}
W_1(s)&=& -G E \left[ \int dt {\rm Pf}_{2s/c}\int \frac{dt'}{|t-t'|}{\mathcal F}_{\rm cm}(t,t')\right]^{\rm on-shell}\,,\nonumber\\
W_2(s)&=& 2G E \left[\int dt {\mathcal F}_{\rm cm}(t)\ln \left(\frac{r_{12}^h(t)}{s}\right) \right]^{\rm on-shell}\,, \nonumber\\
W_{\rm  f-h}&=& 2G E \left[\int dt {\mathcal F}_{\rm cm}(t)\ln \left(f(t)\right) \right]^{\rm on-shell}\,.
\eea
Combining Eqs. \eq{Irtot}, \eq{IrW} yields the following expression for the local part of the radial action
\beq 
\label{Irlocexpression1}
 I_r^{\rm hyp, loc} =  I_r^{\rm hyp, tot} + W_1(s) + W_2(s)  + W_{\rm  f-h}\,.
\eeq
This expression, considered at $O(G^4)$, is the basis of the computation of the present paper. 
In Ref. \cite{Bini:2020hmy} we determined the values of the three terms on the right-hand side (rhs)
 of Eq. \eq{Irlocexpression1} at the
6PN accuracy and at PM orders $G^4$ and $G^5$, see Eqs. (3.37), (3.38), (3.39), (3.40), (3.43), (3.44) in \cite{Bini:2020hmy} (in terms of the orbital parameters $e_r$and $\bar a_r$, in turn related to $p_\infty$ and $j$). 

It is worth to recall here
our  $O(G^4)$ results for $W_1(s)$ and $W_2(s)$, expressed in terms of $j$ and $\pinf$ (or better in terms of $hj$ and $\pinf$, since $(hj)/p_\infty=b/(GM)$):
\bea
\label{W1_W2_6PN}
W^{G^4}_1(s) &=&\frac{\nu^2 GM^2\pi p_\infty^4}{(hj)^3}  \left[\frac{40}{3}+\left(-\frac{317}{105}\nu+\frac{137}{30}\right) p_\infty^2\right.\nonumber\\
& +&  \left(\frac{826619}{60480}-\frac{1444}{945}\nu+\frac{1103}{864}\nu^2\right) p_\infty^4\nonumber\\
&+& \left.2\frac{{\widehat {\mathcal E}}_{\rm 2PN}}{p_\infty} \ln \left(\frac{p_\infty^2 \hat s}{4hj} \right)\right]\,,\nonumber\\ \\
W^{G^4}_2(s) &=&\frac{\nu^2GM^2 \pi p_\infty^4}{(hj)^3} \left[
-\frac{17}{6}+\left(\frac{193}{210}\nu-\frac{4919}{1680}\right) p_\infty^2  \right.\nonumber\\
&+&\left(-\frac{36283}{17280}-\frac{106621}{60480}\nu+\frac{16057}{30240}\nu^2\right)p_\infty^4\nonumber\\
&+&\left. 2\frac{{\widehat {\mathcal E}}_{\rm 2PN}}{p_\infty} \ln \left(\frac{2hj}{\hat sp_\infty}\right)
\right] \,,
\eea
where,   $\hat s=s/(GM)$  and
where
\beq
{\widehat {\mathcal E}}_{\rm 2PN}=\frac{37}{15}p_\infty+\frac{1357}{840}p_\infty^3+\frac{27953}{10080}p_\infty^5+O(p_\infty^7)\,,
\eeq
is the fractionally 2PN-accurate energy loss function ${\widehat {\mathcal E}}$ discussed in its exact form below, see Eq. \eqref{exactE}.

Note that the log in $W^{G^4}_1(s)$ can also be written as $\ln \left(\frac{p_\infty}{2}\frac{s}{2b}\right)$ (and simplifies as $\ln \left(\frac{p_\infty}{2}\right)$ for $s=2b$) whereas the one in $W^{G^4}_2(s)$ becomes 
$\ln \left(\frac{2b}{s}\right)$ (and vanishes for $s=2b$). 

When considering  the $O(G^4)$ level at all orders in the $\pinf$ expansion, two of the various terms on the rhs
of  Eq. \eq{Irlocexpression1} have been  derived: $ I_r^{\rm hyp, tot, G^4}$ has been derived in
Refs. \cite{Dlapa:2021vgp,Bern:2022jvn}, while $W_1(s)$ has been recently obtained in Ref. \cite{Dlapa:2024cje}. The main purpose of the present work is to give
an exact integral expression for $W_2(s)$,  and to compute it to very high order in velocities  (see Eq. \eqref{W20sf} below).

A simplification arises when noticing that,  the 4PM-level gravitational-wave flux being proportional to $\nu^2$, 
and the flexibility factor being (because of Eq. \eq{fnu}) such that $\ln f = O(\nu)$, the last term
on the rhs of Eq. \eq{Irlocexpression1} is of order $W_{\rm  f-h}= O(\nu^3)$. Consequently, the projection
of Eq. \eq{Irlocexpression1} on its $\propto \nu^2$ piece yields
\be \label{Irlocexpressionnu2}
 I_{r, G^4}^{\rm hyp, loc, \nu^2} =  I_{r, G^4}^{\rm hyp, tot, \nu^2} + W_{1, G^4}^{\rm \nu^2}(s) + W_{2, G^4}^{\rm \nu^2}(s)  \,.
\ee
or, equivalently, after multiplying both sides by $\frac{(h j)^3}{G m_1 m_2}$ and using the notation above both
for $ I_{r}^{\rm hyp, tot}$ and $ I_{r}^{\rm hyp, loc}$
\bea 
\label{Irlocexpressionnu1}
\widehat{ I}_{r, 3}^{\rm hyp , loc, \nu^1}(\g)&=& \widehat{ I}_{r, 3}^{\rm hyp , tot, \nu^1}(\g)+ \left[ \frac{(h j)^3 W_{1, G^4}(s)}{G m_1 m_2}  \right]^{\nu^1}\nonumber\\
&+& \left[ \frac{(h j)^3  W_{2, G^4}(s)}{G m_1 m_2}\right]^{\nu^1}\,.
\eea
In other words, for the purpose of computing the gauge-invariant content of the local  4PM dynamics it suffices to 
compute the $\nu^2$ piece of $W_{2, G^4}$. As shown (in a PN-expanded manner) in Refs. \cite{Bini:2020nsb,Bini:2020hmy} the 
$\nu^3 + \nu^4 +\cdots$ parts of $W_{1, G^4}$ and $W_{2, G^4}$ are only needed if one wants to compute
the flexibility factor $f(t)$ (as was done there at the 6PN accuracy). The dependence upon the arbitrary scale $s$ 
cancells when adding $W_1$ and $W_2$.

When considering bound states, the (total) elliptic-motion radial action, $I_r^{\rm ell, tot}$, does not need to be
regularized because it is formally given by a  finite integral over one radial period, say
\be
I_r^{\rm ell, tot}= \oint P_R dR\,.
\ee

However, the only concrete way to compute $I_r^{\rm ell, tot}$ is to work within a PN-expanded (rather
than PM-expanded) framework, so as to be able to define the nonlocal part of the  Hamiltonian, similar to
Eq. \eq{Hnonloc0} (for instance, the nonlocal contribution at order $G^5/c^{11}$ has a 
different structure, as discussed in Ref. \cite{Blanchet:1987wq,Foffa:2011np,Damour:2014jta,Damour:2015isa,Galley:2015kus,Bini:2020hmy}.  

In addition, as emphasized in Ref. \cite{Damour:2014jta,Damour:2015isa}
the nonlocal dynamics does not admit a usual, finite-order $G$-expansion, but rather an infinite-order
expansion in powers of the eccentricity. Such an expansion is non-perturbative in $G$ because the eccentricity is
\be
 e\sim \sqrt{1+ \pinf ^2 j^2}= \sqrt{1+ \frac{(\g^2-1) J^2}{(G m_1 m_2)^2}}
 \ee
 By contrast, the local dynamics defined (in the TF way) by subtracting the nonlocal piece
 of the action from the total action leads to a local Hamiltonian which admits (at each PN order)
 a finite $G$-expansion. We give below an explicit, EOB-based, local-Hamiltonian description of
 the local dynamics. Let us only mention here that the corresponding elliptic-motion local radial action
 \be
I_r^{\rm ell, loc}=\left[ \oint P_R dR \right]^{\rm loc}\,,
\ee
has the following very simple structure \cite{Bini:2020nsb} (in the latter reference the  local action was defined as $\frac1{2\pi} \widehat I_r^{\rm ell , loc}$ so as to factor out all an overall $2\pi$)
\bea
\label{Ir_ell_loc}
{\widehat I}_r^{\rm ell , loc} &=& \frac{I_r^{\rm ell, loc}}{G m_1 m_2}= - 2\pi j +  2 \pi\frac{2 \g^2-1}{\sqrt{1-\g^2}} \nonumber\\
&+&\sum_{n=1,3,5,\cdots}  \frac{ \widehat{ I}_r^{\rm ell , loc}{}_n (\g,\nu) }{(h \,j)^{n}}\,,
\eea
containing only odd powers of $j$, and where each coefficient $\widehat{ I}_r^{\rm ell , loc}{}_n (\g,\nu)$ 
is simply equal to twice its hyperbolic analog:
\be
\widehat{ I}_r^{\rm ell , loc}{}_n (\g,\nu)= 2 \,{\widehat I}_r^{\rm hyp , loc}{}_n (\g,\nu)\,.
\ee
This simple result follows from combining the link between the periastron advance $\Phi= 2\pi+ \delta \Phi$
and the radial action, $\Phi= - \frac{\D I_r^{\rm ell , loc} }{\D J}$, 
the relation (valid for a local Hamiltonian) between the
scattering angle and $\Phi$ \cite{Kalin:2019rwq} (here extended by adding the 
unperturbed angles)
\beq
\Phi(\g, j) = {\rm AC} \left[ (\pi+ \chi(\g, j)) +  (\pi+ \chi(\g, -j)) \right]\,,
\eeq
where AC means \lq\lq Analytic Continuation,"  which formally yields
\beq
 I_r^{\rm ell , loc}(\g,j)=  {\rm AC} \left[  I_r^{\rm hyp , loc}(\g,j) -  I_r^{\rm hyp , loc}(\g,-j) \right]\,.
\eeq
The analytic continuation must be done with respect to the two variables $\g,j$ ($\g$ crossing the value 1 and $j$
changing sign). The second term $ 2 \pi\frac{2 \g^2-1}{\sqrt{1-\g^2}} $ (which was directly obtained in \cite{Bini:2020nsb})
is viewed here as coming from the continuation of the $\frac{\ln j}{\sqrt{\g^2-1}}$ contribution to the hyperbolic radial action as
$ j \to - j$ and $\g^2-1 \to - (\g^2-1)$.

\subsection{Hamiltonian formulation of the local dynamics within EOB theory}

For deriving  physical observables from the 4PM local dynamics it is convenient to translate the
gauge-invariant information contained in $ I_r^{\rm hyp, loc} $ into an explicit Hamiltonian. A useful
way of doing so is to use EOB theory. In  this approach
 the local action (which describes the relative motion in the cm frame) can be put in the form (after
applying a suitable canonical transformation)
\be \label{Sloc}
S_{\rm loc, f}=\int \left( \sum_{i=1}^{3} P_i dQ^i - H_{\rm loc}(Q,P) dt \right)\,,
\ee
where the local Hamiltonian $H_{\rm loc}(Q,P)$ is obtained by solving with respect to
the (real) energy $E$ an EOB mass-shell condition of the general form

\be \label{massshell1}
0=  g^{\mu\nu}_{\rm eff}(Q) P_\mu P_\nu + \mu^2 + {\mathcal Q}^{\rm loc}(Q,P)\,,
\ee
in which the conserved {\it effective} energy 
\be
\Ef= - P_0
\ee
is related to the real energy $E$ via the EOB energy map
\be \label{EvsEf}
E = M \sqrt{1 + 2 \nu \left( \frac{\Ef}{\mu} -1\right) }\,.
\ee
Comparing this relation with Eq. \eq{defh} we see that, when considering scattering states (with $E>M$),
the effective EOB energy is simply related to the asymptotic Lorentz factor $\g$ via \cite{Damour:2016gwp}
\be \label{Efvsg}
\hEf \equiv \frac{\Ef}{\mu} = \g\,.
\ee
On the other hand, the EOB mass-shell condition, Eq. \eq{massshell1}, and the corresponding
local Hamiltonian action,  Eq. \eq{Sloc}, are uniformly applicable to the description of bound states
($E<M$, $\hEf <1$) and scattering states ($E>M$, $\hEf = \g >1$). There arise no subtleties
linked to analytically
continuing the dependence of the Hamiltonian on $\g$ from $\g>1$ to $\g <1$. This lack of
ambiguity in the analytic continuation of the local dynamics is rooted in the fact that
the velocity expansions entering the conservative local action contain only even powers of 
$\pinf$ (or  $v$). Note that, at order $\frac{G^5}{c^{11}}(1+ \frac1{c^2} + \cdots)$, i.e.
5PM and 5.5PN, second-order tail effects give rise to a {\it conservative nonlocal} action (studied in Ref. \cite{Bini:2020nsb})
which is odd in velocities. However, it is purely nonlocal and does not contribute to the corresponding local
action.

A convenient form of the EOB mass-shell condition Eq. \eq{massshell1} is the ``energy gauge," where
the effective metric $g^{\mu\nu}_{\rm eff}$ is taken to be the Schwarzschild metric and where
the additional (higher-than-quadratic in momenta) term ${\mathcal Q}^{\rm loc}(Q,P)$ depends only 
on $P_0= -\g$, and on $GM/R$ (with
$R \equiv | Q^i |$) and admits a PM expansion of the form
\bea 
\label{Qeob}
\frac{{\mathcal Q}^{\rm loc}(Q,P)}{\mu^2}&=& q^{\rm loc}_2(\g,\nu) \left(\frac{GM}R\right)^2+  q^{\rm loc}_3(\g,\nu) \left(\frac{GM}R\right)^3\nonumber\\
&+& q^{\rm loc}_4(\g,\nu) \left(\frac{GM}R\right)^4+ \cdots
\eea
At each PM order $G^n$ (with $n \geq 2$) the local dynamics is fully described by the corresponding
function $q^{\rm loc}_n$. [At order $n=1$ the 1PM dynamics is simply described, in the EOB formalism, by
the (linear part of the) Schwarzschild metric \cite{Damour:2016gwp}.] 

At PM orders $n=2$ and $n=3$, there is
no nonlocal contribution to the dynamics so that $ q^{\rm loc}_2(\g,\nu)$ and $ q^{\rm loc}_3(\g,\nu)$ describe
the full corresponding conservative dynamics. The values of $  q_2(\g,\nu)=q^{\rm loc}_2(\g,\nu)$ and 
$ q_3(\g,\nu) =q^{\rm loc}_3(\g,\nu)$ are in one-to-one correspondence with the corresponding PM
coefficients of the scattering angle, Eq. \eq{chiPM}. The special $\nu$-structure of the scattering angle,
 Eq. \eq{tildechi}, implies a corresponding $\nu$-structure for  $q^{\rm loc}_n$. The choice  
 $ g^{\mu\nu}_{\rm eff}=  g^{\mu\nu}_{\rm Schwarzschild} $ implies that all the $q^{\rm loc}_n(\g,\nu)$'s
 must vanish in the probe limit, i.e. as $\nu \to 0$. It also implies that all the  $q^{\rm loc}_n(\g,\nu)$'s
 must vanish in the nonrelativistic limit $\g \to 1$.
 
 The value of $  q_2(\g,\nu)=q^{\rm loc}_2(\g,\nu)$ is
 \bea
q_2&=&\frac{\nu}{h(h+1)}A_{2,1}(\gamma)\,,\nonumber\\
A_{2,1}(\gamma)&=& 3 (\g-1) (5 \g^2-1)\,.
\eea
The value of $  q_3(\g,\nu)=q^{\rm loc}_3(\g,\nu)$ is
 \bea
q_3&=&\frac{\nu}{h(h+1)}A_{3,1}(\gamma)+ \frac{\nu}{h^2}A_{3,2}(\gamma)\,,\nonumber\\
A_{3,1}(\gamma)&=& -3\frac{(2\g^2-1)(5\g^2-1)}{\g+1}\,.\nonumber\\
A_{3,2}(\gamma)&=& \frac43 \gamma (14\gamma^2+25)\nonumber\\
&+& \frac{8(4\gamma^4-12\gamma^2-3)}{\sqrt{\gamma^2-1}}{\rm arcsinh}\left( \sqrt{\frac{\gamma-1}{2}}\right)\,.\nonumber\\
\eea
Finally, the structure of $q^{\rm loc}_4(\g,\nu)$ reads
\bea \label{q4loc}
q_4^{\rm loc}(\g,\nu)&=&\frac{\nu}{h(h+1)}A_{4,1}(\gamma)+\frac{\nu}{h^2}A_{4,2}(\gamma)\nonumber\\
&+& \frac{\nu}{h^3}A_{4,3}(\gamma)\,.
\eea
Among the three functions of $\g$ entering this structure, two of them are derived from
the lower order results, namely
\begin{widetext}
\bea 
\label{A41A42}
A_{4,1}(\gamma) &=& \frac{(1875\gamma^6-2529\gamma^4+905\gamma^2-59)}{8(\gamma+1)(\gamma^2-1)}\,,\nonumber\\ 
A_{4,2}(\gamma) &=& -\frac{(1347\gamma^5-675\gamma^4+706\gamma^3+270\gamma^2-373\gamma-27)}{12 (\gamma^2-1)}
-\frac{8 (3\gamma^2-1) (4\gamma^4-12\gamma^2-3) }{ (\gamma^2-1)^{3/2}}{\rm arcsinh}\left( \sqrt{\frac{\gamma-1}{2}}\right)\,,\nonumber\\
\eea
\end{widetext}
while the third one, $A_{4,3}(\gamma)$, embodies  genuinely new information about the 4PM
local dynamics. In our previous work \cite{Bini:2020nsb} we have determined the first five   terms in the velocity
expansion of  $A_{4,3}(\gamma)$. We found that it contains both rational coefficients and multiples of $\pi^2$.
Decomposing it as
\be \label{A43}
A_{4,3}(\gamma) = A_{4,3}^{\pi^2}(\gamma) + A_{4,3}^{\rm rem}(\gamma)\,,
\ee
we found
\bea \label{A43pi26PN}
A_{4,3}^{\pi^2}&=&   \pi^2 \left[  -  \frac{41}{32}  
 -  \frac{33601}{6144}  p_\infty^2 \right.\nonumber\\
 &-&  \frac{93031}{12288} p_\infty^4  -  \frac{9733841}{4194304} p_\infty^6 ]\,,
\eea
and  
\bea \label{A43rem6PN}
A_{4,3}^{\rm rem}
&=&   \frac{10}{p_\infty^2} 
+\frac{767}{6}   + \frac{4033}{18}   p_\infty^2 \nonumber\\
&+&  \frac{6514457}{50400}  p_\infty^4  - \frac{6859063}{235200}  p_\infty^6\,.
\eea
Using the results of Sec. X of Ref. \cite{Bini:2020nsb} we know that $A_{4,3}(\gamma) $ is related to the
$\nu^0$ and $\nu^1$ 4PM terms, $\widehat{ I}_{r, 3}^{\rm hyp , loc}= \widehat{ I}_{r, 3, S}^{\rm hyp , loc} + \widehat{ I}_{r, 3}^{\rm hyp , loc, \nu^1}$, in the $G m_1 m_2$-rescaled local action as
\beq 
\label{A43vsIr3}
A_{4,3}(\gamma) = \frac{8}{\pi (\gamma+1)} \left( \widehat{ I}_{r, 3, S}^{\rm hyp , loc}(\g) - \frac{ \widehat{ I}_{r, 3}^{\rm hyp , loc, \nu^1}}{ 2 \nu(\g-1)}\right)\,,
\eeq
where $\widehat{ I}_{r, 3, S}^{\rm hyp , loc}(\g)$ is the Schwarzschild radial action, given by  Eq. \eqref{schw_Ir3},  or equivalently by
\beq
\widehat{ I}_{r, 3, S}^{\rm hyp , loc}(\g)=\pi \left(\frac{35}{4}+\frac{105}{4}p_\infty^2+\frac{1155}{64}p_\infty^4\right)\,.
\eeq
  
This link allows one to deduce from the results of Refs. \cite{Bern:2021yeh,Dlapa:2024cje}
 the exact value of 
$A_{4,3}^{\pi^2}(\gamma)$, and to obtain an exact integral expression for the remaining term 
$ A_{4,3}^{\rm rem}(\gamma)$. Our high-order velocity expansion of $W_2$ will also give the
PN expansion of $ A_{4,3}^{\rm rem}(\gamma)$ to a very high order (see Sec. \ref{secwithW2new} below).

\section{Exact integral expression for $W_2$.}

 As explained above, we only need to compute here the $W_2(s)$ integral defined in Eqs. \eq{Ws}.
 Inserting Eq. \eq{Fttp_Waveform} (taken for $t=t'$) the integral to compute reads
 \be
 W_2(s)=  G E  \left[\int dt {\mathcal F}_{\rm cm}(t)\ln \left( \left(\frac{r_{12}^h(t)}{s}\right)^2 \right) \right]^{\rm on-shell}\,,
 \ee
 with (after replacing the retarded time $u$ by $t$)
 \bea
\label{Ft}
{\mathcal F}_{\rm cm}(t)&=& \frac{1}{16 \pi G}\int d\Omega_{\rm cm} |\dot f_c(t, \theta_{\rm cm}, \phi_{\rm cm})|^2\,,
\eea 
where $\dot f_c(u_{\rm ret}^{\rm cm}, \theta_{\rm cm}, \phi_{\rm cm}) \equiv \frac{df_c(u_{\rm ret}^{\rm cm}, \theta_{\rm cm}, \phi_{\rm cm})}{du_{\rm ret}^{\rm cm}}$
denotes the (complex) gravitational news function at retarded time $u$. 
The news function starts at the PM order  $\dot f_c(u) = O(G^2)$ so that ${\mathcal F}_{\rm cm}= O(G^3)$ and 
 $W_2(s)=   O(G^4)$.

Before tackling the computation of the integral $W_2$, let us remark that while $W_1$ involves a nonlocal-in-time
kernel $\sim \frac1{|t-t'| }$ acting on a time-split energy flux ${\mathcal F}_{\rm cm}(t,t')$,  $W_2$ involves a local-in-time kernel $\sim \delta(t-t')$ diagonalizing ${\mathcal F}_{\rm cm}(t,t')$ into  ${\mathcal F}_{\rm cm}(t)$. 
The situation would be reversed when going to Fourier space,
i.e., when replacing $f_c(t, \theta, \phi)$ 
by  $\widehat f_c(\omega, \theta, \phi)=\int dt e^{i\omega t} f_c(t, \theta, \phi)$.
 In Fourier space,  the integral $W_1$ diagonalizes to 
\beq
W_1(s)=GE\int \frac{d\omega}{2\pi} \ln (4s^2 e^{2\gamma}\omega^2)   {\mathfrak F}_{\rm cm}(\omega)\,, 
\eeq
where 
\beq
 {\mathfrak F}_{\rm cm}(\omega)= \frac{1}{16 \pi G}\int d\Omega_{\rm cm}  |- i \omega \widehat f_c(\omega, \theta_{\rm cm}, \phi_{\rm cm})|^2\,.
\eeq
is the frequency spectrum of the emitted gravitational-wave radiation. 
[Note in passing that $  {\mathfrak F}_{\rm cm}(\omega)$ is not the Fourier transform of the function
 $ {\mathcal F}_{\rm cm}(t)$.] 
 
 By contrast, the Fourier-space version of the
 integral $W_2$ involves the naturally defined frequency-split version $  {\mathfrak F}_{\rm cm}(\omega, \omega')$
 of $ {\mathfrak F}_{\rm cm}(\omega)$, together with the Fourier transform of 
 $\ln \left( \left(\frac{r_{12}^h(t)}{s}\right)^2 \right)$. We found more convenient to compute $W_2$
 in the time domain.

 When restricting ourselves
to the $G^4$ level, as we do here,  we can approximate the news function by its leading-order PM value.
This corresponds to the leading-order (time-dependent)
  brehmsstrahlung asymptotic waveform first computed (as an explicit function of retarded time) in the frame of one particle by Ref. \cite{kov-tho4}, and recently recomputed by efficient, quantum-based approaches in Refs. \cite{Kosower:2018adc,Mougiakakos:2021ckm,Jakobsen:2021smu,DeAngelis:2023lvf} 

As ${\mathcal F}_{\rm cm}(t) = O(G^3)$, we can replace $\left(r_{12}^h(t)\right)^2$ in the integrand
  by its  zeroth-order PM value, i.e. the square of the cm-frame distance between the two particles moving
  along two straight worldlines, say\footnote{We denote the two bodies either as   1 and   2 or   A and B when convenient.}
\bea
\label{worldlines}
z^\mu_A(\tau_A)&=&z_A^\mu(0)+u_A^\mu \tau_A + O(G)\,,\nonumber\\
z_B^\mu(\tau_B)&=& z_B^\mu(0)+u_B^\mu\tau_B + O(G)\,,
\eea
where  $\tau_A$ and $\tau_B$ are the (Minkowski)  proper times along the worldlines,
and where
\be
z_A^\mu(0)- z_B^\mu(0)= b^\mu
\ee
is the vectorial impact parameter, orthogonal to both incoming four-velocities $u_A^\mu$ and $u_B^\mu$, 
and of magnitude $b = |  b^\mu|= \sqrt{ \eta_{\mu \nu} b^\mu b^\nu}$. 
  
In order to compute $W_2$ we need the explicit cm-frame values of 
\be
r_{12}^2(t_{\rm cm})= |z^\mu_A(\tau_A)- z_B^\mu(\tau_B)|^2_{\rm cm}
\ee
and of the differential cm energy flux per unit retarded time and per unit solid angle
\be \label{cm_diff_flux0}
\frac{d {\mathcal F}^{\rm rad}_{\rm cm}}{d\Omega_{\rm cm}}(u_{\rm ret}^{\rm cm},\Omega_{\rm cm} )=\frac{dE^{\rm rad}_{\rm cm}}{du_{\rm ret}^{\rm cm}d\Omega_{\rm cm}} (u_{\rm ret}^{\rm cm},\Omega_{\rm cm} )\,.
\ee
Here, $t_{\rm cm}$ denotes the coordinate time in the cm frame (with origin $t_{\rm cm}=0$
corresponding to the cm instant of closest approach of the two bodies), while 
$u_{\rm ret}^{\rm cm} = t_{\rm cm} - r +O(G)$ is the corresponding lightcone coordinate used
at future null infinity to parametrize the waveform $ f_c(u_{\rm ret}^{\rm cm}, \Omega_{\rm cm})$.
After inserting $\frac{d {\mathcal F}^{\rm rad}_{\rm cm}}{d\Omega_{\rm cm}}(u_{\rm ret}^{\rm cm},\Omega_{\rm cm} )$ in the $W_2$ integral one replaces $u_{\rm ret}^{\rm cm}$ by the dynamical time variable $t_{\rm cm}$
entering $r_{12}^2(t_{\rm cm})$. In other words, we have
\bea
\label{W2cmdef}
W_2(s)&=&GE \int dt_{\rm cm} d\Omega_{\rm cm}\ln \left(\frac{r_{12}^2(t_{\rm cm})}{s^2}\right) \times \nonumber\\
&&\left[\frac{dE^{  \rm rad}_{\rm cm}  }{du_{\rm ret}^{\rm cm} d\Omega_{\rm cm} } \right]^{u_{\rm cm}=t_{\rm cm}}  \,,
\eea

Let us give convenient explicit expressions for the building blocks of the integrand defining $W_2$.

First, the value of $r_{12}^2(t_{\rm cm})$ is obtained from
\be
| z^\mu_A(\tau_A)- z^\mu_B(\tau_B) |^2
\ee
by imposing that the two proper times $\tau_A$ and $\tau_B$ correspond to the
cm coordinate time $t_{\rm cm}$. This yields
\beq
t_{\rm cm} = -U\cdot  z_A(\tau_A) = - U\cdot  z_B(\tau_B)\,,
\eeq
where 
\bea
U\equiv u_{\rm cm}&=&\frac{m_A u_A +m_B u_B}{E}
\eea
is the unit time vector of the cm frame. This implies the following explicit
relation between $\tau_A$,  $\tau_B$,  and $t_{\rm cm}$
\beq
t_{\rm cm}=\frac{E_A}{m_A}\tau_A = \frac{E_B}{m_B}\tau_B \,,
\eeq
where $E_A$ and $E_B$ are the cm energies of particles $A$ and $B$, given by
\beq
\frac{E_A}{m_A}=\frac{m_A+\gamma m_B}{E}\,,\qquad \frac{E_B}{m_B}=\frac{m_B+\gamma m_A}{E}\,,
\eeq
where, as above, $E= E_A + E_B$ is the total cm energy.

Combining these relations yields
\bea
r_{\rm cm}^2&=&b^2+\left( \frac{m_Am_B}{E_AE_B}\right)^2\gamma^2v^2 t_{\rm cm}^2\nonumber\\
&=& b^2(1+T_{\rm cm}^2)\,,
\eea
where we defined the dimensionless cm time $T_{\rm cm}$ as
\beq
T_{\rm cm} \equiv \frac{\gamma v}{b}\frac{m_A m_B}{E_A E_B} t_{\rm cm}\,.
\eeq

The second building block is the cm-frame based differential energy flux Eq. \eq{cm_diff_flux0}. To obtain its explicit
expression starting from the explicit waveform expressions of Refs. \cite{kov-tho4,Jakobsen:2021smu} 
(which are
given in the rest frame of particle $A$), one needs the transformation formula relating the $A$-rest-frame
differential flux
\beq
\label{A_diff_flux}
\frac{dE_A}{du_{\rm ret}^A d\Omega_A}(u_{\rm ret}^A,\Omega_A)du_{\rm ret}^A d\Omega_A
\eeq
to its cm-frame analog
\beq
\label{cm_diff_flux}
\frac{dE_{\rm cm}}{du_{\rm ret}^{\rm cm}d\Omega_{\rm cm}}(u_{\rm ret}^{\rm cm},\Omega_{\rm cm} ) du_{\rm ret}^{\rm cm}d\Omega_{\rm cm}\,.
\eeq
Alternatively, one can start from the general-frame time-domain waveform of Ref. \cite{DeAngelis:2023lvf}, 
and apply them to the cm frame.

When starting from the $A$-frame complex waveform $f^A_c(u_A, \theta_A, \phi_A)$, where
$u_A\equiv  u_{\rm ret}^A=\tau_A- |x-z_A(0)| $ and its associated differential flux
\bea
\label{Ftbis}
\frac{dE_A}{du_A d\Omega_A}&=& \frac{1}{16 \pi G} \bigg|  \frac{d }{du_A}f^A_c(u_A, \theta_A, \phi_A)\bigg|^2\,,
\eea 
we have the following simple transformation law for the differential fluxes
\begin{widetext}
\beq
\label{flux_cm}
\frac{dE_{\rm cm}}{du_{\rm ret}^{\rm cm}d\Omega_{\rm cm}}(u_{\rm ret}^{\rm cm},\Omega_{\rm cm} ) du_{\rm ret}^{\rm cm}d\Omega_{\rm cm}=f_{\rm cm} \frac{dE_A}{du_{\rm ret}^A d\Omega_A}(u_{\rm ret}^A,\Omega_A)du_{\rm ret}^A d\Omega_A\,, 
\eeq
\end{widetext}
involving the following Doppler factor between the $A$-frame and the cm frame 
\be \label{Doppler}
f_{\rm cm}= \gamma_{\rm cm}(1-\alpha_A v_{\rm cm}) = \frac{1}{\gamma_{\rm cm} (1+\alpha_{\rm cm} v_{\rm cm} )}\,.
\ee
Here, $ v_{\rm cm}$ denotes the relative velocity between particle $A$ and the  cm frame (with $\gamma_{\rm cm}=
1/\sqrt{1-  v_{\rm cm}^2}$), while $\alpha_A = \cos \theta_A$ and $\alpha_{\rm cm}=  \cos \theta_{\rm cm}$,
with $\theta_A$ and $\theta_{\rm cm}$ being the angles of gravitational-wave emission respectively recorded in the $A$-frame
and the cm frame. See Appendix \ref{AppFrames} for a derivation of this transformation law (and for the associated
transformation laws of $E$, $u$ and $\alpha = \cos \theta$).

As the integrand for $W_2$ contains the logarithm of the squared cm distance expressed as a function of $t_{\rm cm} = u_{\rm cm} + {\rm cst}$,
while the $A$-frame waveform is expressed as a function of $u_A$, we  need the transformation law between $u_A$
and $ u_{\rm cm}$. It is 
\be \label{uAvsucm}
u_A = f_{\rm cm}(\alpha_A)  u_{\rm cm} - \beta_A  b_A\,,
\ee
where $\beta_A = \sin \theta_A \cos \phi_A$ and where $b_A=\frac{E_B}{E}b$ is the relative
impact parameter between particle $A$ and the spatial origin of the cm system. 

This finally  leads to the following explicit expression (involving the square of the
Doppler factor) of $W_2$ in terms of the $A$-frame waveform
\bea 
\label{W2Aframe}
W_2(s)&&=GE \int d u_{\rm cm} d\Omega_A f_{\rm cm}^2(\alpha_A)  \ln \left(\frac{r_{12}^2\big|_{t_{\rm cm}= u_{\rm cm}}}{s^2}\right) \times \nonumber\\
&&\left[  \frac{dE_A}{du_A d\Omega_A}(u_A,\Omega_A)\right]^{u_A = f_{\rm cm}(\alpha_A)  u_{\rm cm} - \beta_A  b_A}\,,\qquad
\eea
in which the integrand has to be re-expressed as a function of $u_{\rm cm}$ ($ = t_{\rm cm}$) and of the
$A$-frame angles $\theta_A$, $ \phi_A$. This leads to replacing everywhere in the news function  
$\frac{d f^A_c(u_A, \theta_A, \phi_A)}{du_A}$ the variable $u_A$ by the rhs of Eq. \eq{uAvsucm}.

The original cm-frame definition, Eq. \eqref{W2cmdef},  of $W_2$ makes it clear that $W_2$ is a {\it symmetric} function
of the two masses, $m_A$ and $m_B$ (and therefore a function of $M= m_A+m_B$ and of the
symmetric mass ratio $\nu$. By contrast, the reformulation, Eq. \eq{W2Aframe}, of $W_2$ in terms of 
$A$-frame quantities treats in a very dissymetric manner body $A$ and body $B$. For instance, the
Doppler factor $f_{\rm cm}(\alpha_A)= \gamma_{\rm cm}(1-\alpha_A v_{\rm cm}) $ involves 
the emission angle in the $A$-frame as well as the relative velocity $v_{\rm cm}$ between body $A$ and the cm frame.
In addition, the transformation between $u_A$ and $u_{\rm cm}$ is also quite dissymetric.
As a check of Eq. \eq{W2Aframe}, we have verified that the velocity expansion of the integrand in Eq. \eq{W2Aframe}
is indeed a symmetric function of $m_A$ and $m_B$ which does not involve odd powers of  $\Delta = \frac{m_A-m_B}{M}$.

\section{Computation of the contribution of $W_2$ to the local action}
\label{secwithW2new}

As just recalled, $W_2(s)$ is a symmetric function of $m_A$ and $m_B$. In addition, being proportional to
the radiated flux, it contains an overall factor $ (m_A m_B)^2 = \nu^2 M^4$. Therefore, its $\nu$ dependence
is $\propto \nu^2 (1+ \nu+ \nu^2 + \cdots)$. For the reasons recalled in Eq. \eq{Irlocexpressionnu2} above, the terms of
order $\nu^3$ and higher in $W_2$, together with similar contributions contained in $W_1$ (which is also
 $\propto \nu^2 (1+ \nu+ \nu^2 + \cdots)$) do not contribute to the local action, but only contribute to the determination
 of the flexibility factor $f(t)= 1+ O(\nu)$ entering the definition of the nonlocal part of the action.
 
 This remark allows us to simplify our task. It suffices to the determine  the $\nu^2$ part of $W_2$,
 and the latter can be obtained by taking the limit, $m_B/m_A \to 0$, in the computation of $W_2$. This drastically simplifies
 the exact expression Eq. \eq{W2Aframe} on three aspects: (i) as $v_{\rm cm} \to 0$ when $m_B/m_A \to 0$ 
 the Doppler factor $f_{\rm cm}$, Eq. \eq{Doppler}, can be replaced by 1; (ii) the angle-dependent transformation
 \eq{uAvsucm} becomes trivial, $u_A = u_{\rm cm}$ because $b_A \to 0$; and (iii) the squared distance
 $r_{12}^2(t_{\rm cm})$ also simplifies to
 \beq
r_{12}^2(u_{\rm cm})=b^2+\frac{u_{\rm cm}^2}{\gamma^2}\,.
\eeq
  Finally, we can directly express the useful $\nu^2$ part of $W_2$ in terms of  $A$-frame quantities. Introducing
  the convenient rescaled $A$-frame time variable (in which we use the identification $u_A= \tau_A$ derived from the identification $u_{\rm cm} = t_{\rm cm} $
  explained above)
  \be
  T_A \equiv \frac{\gamma v}{b} u_A = \frac{\gamma v}{b} \tau_A\,,
  \ee
  which naturally enters the $A$-frame waveform (together with the other rescaled time $T_B$ related to
  $T_A$ by an angle-dependent relation, see Eq. \eqref{TB_vs_TA}), we finally get
  \bea 
\label{W2final}
W_2^{\nu^2}(s)&=&GM  
\int dT_A \sin \theta_A d\theta_A d\phi_A  \times\nonumber\\
&&\frac{dE^{\rm rad}_A}{ dT_A d\Omega_A}(T_A,\theta_A, \phi_A) \ln \left(\frac{b^2(1+\frac{T_A^2}{\gamma^2})}{s^2}\right)\,,\qquad
\eea
where we also replaced $E$ by $M$ because we are considering the $\nu^2$ part of $W_2$.

The $O(G^2)$, $A$-frame, waveform,  $f^A_c(u_A, \theta_A, \phi_A)$ is a rational function of four variables:
$T_A$, $T_B$, $\ell_A \equiv \sqrt{1+T_A^2}$ and $\ell_B \equiv\sqrt{1+T_B^2}$, where
\be
T_A=\frac{\gamma v }{b}\left(\tau_A-|{\mathbf x}-{\mathbf z}_A(0) | \right)=\frac{\gamma v}{b}u_{{\rm ret}}^A\,,
\ee
and
\be
T_B=\frac{\gamma v }{b}\left(\tau_B-|{\mathbf x}-{\mathbf z}_B(0) | \right)=\frac{\gamma v}{b}u_{{\rm ret}}^B\,.
\ee
The link between $T_B$ and $T_A$ reads
\beq
T_B=\frac{T_A+\gamma v \beta_A}{\gamma (1-v \alpha_A)}\,,
\eeq
where, we recall,  
\be
\alpha_A=\cos\theta_A\,, \qquad \qquad \beta_A=\sin\theta_A \cos \phi_A\,.
\ee
In addition, $f^A_c(u_A, \theta_A, \phi_A)$ contains the spurious denonimator
\be
S^2 = \frac{T_A^2 - 2 \g T_A T_B + T_B^2}{\g^2-1}-1\,,
\ee
which vanishes for some (real) angles and time, without, however, corresponding to any
singularity of the waveform. The recent new derivation of the  $O(G^2)$  (general-frame) waveform
\cite{DeAngelis:2023lvf} has shown how to explicitly factor out the vanishing part of the denominator
\beq
S^2= -\frac{\left(\g + T_A T_B - \ell_A  \ell_B  \right)  \left(\g + T_A T_B  +\ell_A  \ell_B  \right)}{\g^2-1} \,.
\eeq

The corresponding radiative flux 
\beq
\label{Ftter}
\frac{dE_A}{dT_A d\Omega_A}(T_A, \Omega_A)=\frac{1}{16 \pi G} \frac{\gamma v}{b}\bigg|\frac{d f^A_c(T_A, \theta_A, \phi_A)}{dT_A}\bigg|^2\,,\qquad
\eeq 
is a rather complicated function of $T_A, \theta_A, \phi_A$. In addition to the denonimators 
 $\ell_A \equiv \sqrt{1+T_A^2}$, $\ell_B \equiv\sqrt{1+T_B^2}$ and $S^2$ (or only $\g + T_A T_B  +\ell_A  \ell_B $),
 the integrand of $W_2^{\nu^2}(s)$ also contains the logarithm  
 $ \ln \left(\frac{b^2}{s^2} (1+\frac{T_A^2}{\gamma^2})\right)$ which involves the new time function
 $(1+\frac{T_A^2}{\gamma^2})$.

We did not succeed in computing the exact value of the integral in Eq. \eq{W2final}. However, we could compute
its velocity expansion up to the thirtieth order. We give some details of the way we performed and checked the
velocity expansion in Appendix \ref{W2_details}. 
As the crucial 
$ \ln \left(\frac{b^2}{s^2} (1+\frac{T_A^2}{\gamma^2})\right)$ factor in the  $W_2^{\nu^2}(s)$ 
integral does not depend on angles one can first integrate over angles, i.e. first compute the instantaneous 
radiated power
\beq
\label{dEdT}
\frac{dE^{\rm rad}_A}{dT_A }(T_A)= \frac{1}{16 \pi G}  \frac{\gamma v}{b}\int  \sin \theta_A d\theta_A d\phi_A \bigg|\frac{d f^A_c(T_A, \theta_A, \phi_A)}{dT_A}\bigg|^2\,.\qquad
\eeq 
 
The dependence of $W_2^{\nu^2}(s)$ on the scale $s$ 
is given by the $A$-frame time-integrated radiated energy 
\bea
E^{\rm rad}_A&=&\int dT_A d\Omega_A\frac{dE^{\rm rad}_A}{ dT_A d\Omega_A}(T_A,\theta_A, \phi_A)\qquad\nonumber\\
&=& \int dT_A \frac{dE^{\rm rad}_A}{dT_A }(T_A)\,,
\eea
with $d\Omega_A= \sin \theta_A d\theta_A d\phi_A$ and
\beq
E^{\rm rad}_A=\frac{\nu^2G^3 M^4}{b^3}\pi \widehat {\mathcal E}(\g)\,,
\eeq
namely
\beq
W_2^{\nu^2}(s)=- 2 \pi \frac{\nu^2G^4 M^5  }{b^3}\widehat {\mathcal E}(\g) \ln \frac{s}{2 b} + W_2^{\nu^2}(s= 2 b)\,.  
\eeq

The exact value of $\widehat {\mathcal E}$ as a function of $\g$
 has been first obtained in Refs. \cite{Herrmann:2021lqe,Herrmann:2021tct} and can be written
 (in the notation of \cite{Bern:2021yeh}) as
\bea
\label{exactE}
\widehat {\mathcal E}(\g)=\frac{{\mathcal M}_4^t}{4\sqrt{\gamma^2-1}}\,,
\eea
with
\bea
{\mathcal M}_4^{\rm t}&=& r_1 +r_2 \ln \left(\frac{\gamma+1 }{2}\right)+r_3 \frac{{\rm arccosh}(\gamma )}{\sqrt{\gamma^2-1}}\,,\qquad
\eea
where
\bea
r_1&=&\frac{1}{12 (\gamma ^2-1)}\left(210 \gamma ^6-552 \gamma ^5+339 \gamma ^4\right.\nonumber\\
&-&\left. 912 \gamma ^3+3148 \gamma ^2-3336 \gamma +1151\right)\,, \nonumber\\
r_2&=& -\frac{35 \gamma ^4}{2}-30 \gamma ^3+75 \gamma ^2-38 \gamma +\frac{5}{2}\,,\nonumber\\
r_3&=& \frac{\gamma  \left(2 \gamma ^2-3\right) \left(35 \gamma ^4-30 \gamma ^2+11\right) }{4 (\gamma ^2-1)}\,.
\eea
We used this exact result as a check on our velocity-expanded computation of $W_2(s)$.
 
Some details of our computation of $W_2^{\nu^2}(s)$ are given in Appendix \ref{W2_details}. We performed two
independent computations. One based on the $A$-frame waveform of  Ref. \cite{kov-tho4},
and one based on the $A$-frame waveform  of  Ref. \cite{Jakobsen:2021smu}.
In addition, we used the recent spurious-pole-free, general-frame, result of Ref. \cite{DeAngelis:2023lvf} as a further check on
our computation. After integration over the angles, the velocity expansion of the instantaneous radiative flux
has the following structure
\beq
\frac{dE^{\rm rad}_A}{dT_A }(T_A) = \sum_{n\geq 1} v^n \sum_{m\geq1} \frac{P_{n,m}(T_A)}{(1+T_A^2)^m} \,,
\eeq
where $P_{n,m}(T_A)$ is a {\it polynomial} in $T_A$ (whose degree increases linearly with the velocity order $n$).
As said above, we checked the correctness of our high $n$ terms by first computing the integral of $\frac{dE^{\rm rad}_A}{dT_A }(T_A)$
over $T_A$ and comparing with the exact result, Eq. \eqref{exactE}.

To compute  $W_2^{\nu^2}(s=2b)$  one must add the logarithmic factor 
 $ \ln \left(\frac14 (1+\frac{T_A^2}{\gamma^2})\right)$. It is first expanded in powers of the
 velocity using $\frac1{\g^2} = 1-v^2$. Besides a series of powers of $1+T_A^2$ this adds the
 logarithmic factor  $ \ln \left(\frac14 (1+T_A^2)\right)$ which is replaced by
 \beq
 \ln \left(\frac14 (1+T_A^2)\right) \to \frac1{\epsilon} \left(  \left(\frac{1+T_A^2}{4}\right)^\epsilon -1 \right) 
 \eeq
 before taking the limit $\epsilon \to 0$.
 
Our final result (new with this work) reads 
\beq
W_{2}^{\nu^2}(s=2b; v)=\left( \frac{GM}{c^2b}\right)^3 \frac{GM^2}{c} W_{2\rm resc}^{\nu^2}(s=2b; v)\,,
\eeq
with
\bea
\label{W20sf}
W_{2\rm resc}^{\nu^2}(s=2b; v)&=&-\nu^2 \pi\left( \frac{17}{6} v  + \frac{2433}{560} v^3 + \frac{913753}{120960} v^5\right.\nonumber\\ 
&+&\frac{226043537}{21288960} v^7 
+ \frac{36845202749}{2767564800} v^9  \nonumber\\
&+&  \frac{21807881107}{1383782400} v^{11} + \frac{879414009003}{48791142400} v^{13}\nonumber\\ 
&+& \frac{21519069553005617}{1067940524851200} v^{15}\nonumber\\
&+& \frac{85221282850631693}{3844585889464320} v^{17}\nonumber\\
&+&  
 \frac{42609586913972092373}{1768509509153587200} v^{19} \nonumber\\
&+& \frac{
 2968979222698685033}{114432968239349760} v^{21}\nonumber\\  
&+&  
\frac{17263645320041062089727}{622515347222062694400} v^{23}\nonumber\\  
&+& \frac{
 727841022305017850589739}{24704030095022909030400} v^{25}\nonumber\\ 
&+&  \frac{
 56643234190964754991093211}{1818834215746061677363200} v^{27} \nonumber\\
&+& \left.  \frac{
 207364235618758023848016841}{6326379880855866703872000} v^{29}
\right)\nonumber\\
&+& O(v^{31})\,,
\eea
or, equivalently, in terms of $p_\infty$ instead of $v$:
\bea
\label{W20sfpinf}
W_{2\rm resc}^{\nu^2}(s=2b; \pinf)&=&
\nu^2 \pi \left(- \frac{17}{6} p_\infty - \frac{4919}{1680 }p_\infty^3 -  
 \frac{36283}{17280}p_\infty^5 \right.\nonumber\\ 
&+&  
 \frac{7144381}{7096320} p_\infty^7  - 
 \frac{72418573}{153753600} p_\infty^9  \nonumber\\ 
&+&   
 \frac{1259235811}{11070259200}  p_\infty^{11}  + 
 \frac{1971149651}{12666931200}  p_\infty^{13}\nonumber\\
&-&  
 \frac{70696880111227}{188460092620800} p_\infty^{15} \nonumber\\
&+& \frac{1085125617350279}{1922292944732160} p_\infty^{17}\nonumber\\
&-&  \frac{2461760980710763}{3349449827942400} p_\infty^{19} \nonumber\\
&+&  \frac{11586658106418094543}{12969069733792972800} p_\infty^{21}  \nonumber\\
&-&  \frac{216662429663169083771}{207505115740687564800} p_\infty^{23}  \nonumber\\
&+&  \frac{11636540057399489827871}{9778678579279901491200} p_\infty^{25}  \nonumber\\
&-&  \frac{1231409296585844895956779}{923852300061491645644800} p_\infty^{27} \nonumber\\
&+&\left.\frac{336216738115283166236999}{228066986300446605312000} p_\infty^{29} 
 \right) \nonumber\\
& +& O(p_\infty^{31})\,,
\eea
The first  three terms in these expressions agree with the $\nu^2$ part of our previous 6PN-accurate (2PN fractional accuracy)
computation of $W_2$ in Ref. \cite{Bini:2020hmy} (Eq. (4.5) there), as recalled in Eqs. \eqref{W1_W2_6PN} above.  

The latter expansion is also provided in electronic form in an ancillary file to this paper.

\section{Computation of the 4PM local hyperbolic action } \label{4PMradialaction}

Having obtained a high-order result for $W_2$, we can now insert it in  Eq. \eq{Irlocexpressionnu1}, together
with the value of $ I_{r, G^4}^{\rm hyp, tot, \nu^2}$ computed in  \cite{Bern:2021yeh,Dlapa:2021vgp} and of $W_1^{G^4}$ computed
in Ref. \cite{Dlapa:2024cje}, to get the 4PM local hyperbolic radial action with a corresponding high PN accuracy.

In terms of the coefficient $\widehat{ I}_{r, 3}^{\rm hyp , loc, \nu^1}(\g)$ of $\frac1{(hj)^3}$ in the
$G m_1 m_2$-rescaled hyperbolic radial action defined in Eq. \eqref{Ir_hyp_loc} we get 
\bea
\label{I3_pi2andrem}
\widehat{ I}_{r, 3}^{\rm hyp , loc, \nu^1}(\pinf) = \pi \nu \left(  i_{r,3}^{\rm hyp,  \pi^2} + i_{r,3}^{\rm hyp , rem}  \right)
\eea
which we separated into two parts: a part whose PN expansion is proportional to $\pi^2$ (after having factored
out an overall factor $\pi \nu$), and a remaining part whose PN expansion has rational coefficients.

The part $ i_{r,3}^{\pi^2}$ was first computed to 6PN accuracy in Ref. \cite{Bini:2020nsb}, and can now 
 (as both $W_1$ and $W_2$ only contribute to the rational part) be
exactly obtained from Refs. \cite{Bern:2021yeh,Dlapa:2024cje}.  It reads in the notation of Ref. \cite{Bern:2021yeh},
\beq
 i_{r,3}^{\pi^2}= - \frac18 \pinf^2{\mathcal M_4}^{\pi^2}\,,
 \eeq
 where
\bea \label{M4pi2}
{\mathcal M}_4^{\pi^2}&=& r_4\pi^2 + r_5 K\left(\frac{\gamma-1}{\gamma+1} \right) E\left(\frac{\gamma-1}{\gamma+1}\right) \nonumber\\
&+& r_6 K^2\left(\frac{\gamma-1}{\gamma+1}\right) + r_7 E^2\left(\frac{\gamma-1}{\gamma+1}\right)\,,
\eea
with
\bea \label{coefM4pi2}
r_4&=& - \frac{16}{3}  (2 + 15\gamma^2) + \frac43 \gamma (19 + 15\gamma^2) \nonumber\\
&+& 
 \frac13 (20 + 111\gamma^2 + 30\gamma^4 - 25\gamma^6)\,,\nonumber\\
r_5&=& \frac{ -1183 - 2929\gamma - 2660\gamma^2 - 
 1200\gamma^3 }{ 2 (\gamma^2-1 ) }\,, \nonumber\\
r_6&=& \frac{(834 + 2095\gamma + 1200\gamma^2)}{ 2 (\gamma^2-1 ) }\,,\nonumber\\
r_7&=& \frac{7 (169 + 380\gamma^2)}{ 4 (\gamma-1) }\,.
\eea

The velocity expansion of $ i_{r,3}^{\pi^2}$ with the same accuracy as our new results read 
\bea
i_{r,3}^{{\rm hyp},\pi^2}
&=&\pi^2\left[ \frac{41}{128} p_\infty^2   +  \frac{33601}{24576} p_\infty ^4  +  \frac{93031}{49152}p_\infty^6\right.\nonumber\\   
&+&  
 \frac{9733841}{16777216 } p_\infty^8 +  \frac{303760055}{805306368} p_\infty^{10}\nonumber\\
&-&  
 \frac{15380987189}{51539607552} p_\infty^{12}  + \frac{16541063287}{68719476736} p_\infty^{14}  \nonumber\\
&-& \frac{27948129277267}{140737488355328} p_\infty^{16} \nonumber\\
&+& \frac{752705766645767}{4503599627370496} p_\infty^{18}\nonumber\\
&-& \frac{123692590730776925}{864691128455135232}p_\infty^{20}\nonumber\\
&+&  \frac{107367805309183991}{864691128455135232} p_\infty^{22}   \nonumber\\
&-& \frac{193147770737983523995}{1770887431076116955136} p_\infty^{24}  \nonumber\\
&+&  \frac{2742100258254883550063}{28334198897217871282176} p_\infty^{26}  \nonumber\\
&-&  \frac{157075421780003581656641}{1813388729421943762059264} p_\infty^{28}   \nonumber\\
&+&  \frac{566610437031857627233835}{7253554917687775048237056} p_\infty^{30}   \nonumber\\
&-&  \frac{1404532330880150975098811995}{19807040628566084398385987584} p_\infty^{32}  \nonumber\\
&+& \left. O\left(p_\infty^{34}\right)\right]\,, 
\eea
The newest contribution of our present work is the following high-velocity expansion of the remaining (rational)
contribution $ i_{r,3}^{\rm hyp , rem} $ in Eq. \eqref{I3_pi2andrem}. We get
\bea
i_{r,3}^{{\rm hyp, rem}}
&=&
 - \frac52   -  \frac{557}{24} p_\infty^2  -  \frac{4601}{144}p_\infty^4  
-\frac{3978707}{201600} p_\infty^6\nonumber\\  
&+&  \frac{5058313}{940800}p_\infty^8  -  
 \frac{2080474861}{223534080}p_\infty^{10}\nonumber\\ 
&+&  
 \frac{7525026057359}{1278614937600}p_\infty^{12}  -  
 \frac{667875191241}{205209804800} p_\infty^{14}\nonumber\\ 
&+&\frac{1586485996991}{1677618708480}  p_\infty^{16}\nonumber\\ 
&+&  \frac{5852773642418587207}{4673094148643880960}  p_\infty^{18}\nonumber\\
&-&  \frac{102439214242834126523}{29596262941411246080}p_\infty^{20} \nonumber\\
&+&  \frac{356424820119856646321}{61883095241132605440} p_\infty^{22}   \nonumber\\
&-& \frac{10274871541456082531814863}{1252513847680523934105600} p_\infty^{24}   \nonumber\\
&+& \frac{1357001641252059026788231699}{125251384768052393410560000} p_\infty^{26}  \nonumber\\
&-& \frac{10091085308176975110532965097}{737382212581842286018560000} p_\infty^{28}   \nonumber\\
&+&  \frac{2367483429515771992293609237895631}{141066526418837018579809861632000} p_\infty^{30}  \nonumber\\
&-& \frac{1762369308619801501266014975258153741}{87461246379678951519482114211840000} p_\infty^{32}   \nonumber\\
&+& O\left(p_\infty^{34}\right)
\,.
\eea
The corresponding 4PM-level contribution to the rescaled scattering angle 
$\tilde \chi_4^{\rm loc}(p_\infty)$ (in the notation of Eq. (8.4) in Ref. \cite{Bini:2020nsb}) 
reads
\beq
\tilde \chi_4^{\rm loc}= \frac{3}{2}  I_{r, 3}^{\rm hyp , loc, \nu^1}= \frac32 \pi \nu \left(  i_{r,3}^{\rm hyp,  \pi^2} + i_{r,3}^{\rm hyp , rem}  \right)\,,
\eeq
while the corresponding term in the {\it bound state} local radial action is
\beq
 I_{r, 3}^{\rm ell , loc, \nu^1}= 2 \,  I_{r, 3}^{\rm hyp , loc, \nu^1}\,.
\eeq


\section{Hamiltonian transcription of the local 4PM radial action within the EOB formalism}

As shown in Sec. \ref{secwithW2new} above our new result can be straightforwardly translated into the 4PM
contribution $q^{\rm loc}_4(\g,\nu) \left(\frac{GM}R\right)^4$ to the additional
term ${\mathcal Q}^{\rm loc}(Q,P)$ in the energy-gauge EOB
mass-shell condition Eq. \eq{massshell1}. More precisely, $q^{\rm loc}_4(\g,\nu)$, as displayed in Eq. \eq{q4loc},
is made of three terms parametrized by the three functions of $\g$, $A_{4,1}(\gamma)$,  $A_{4,2}(\gamma)$,
and  $A_{4,3}(\gamma)$. The first two functions  $A_{4,1}(\gamma)$, and $A_{4,2}(\gamma)$ (which are
determined by the conservative 2PM and 3PM dynamics) were displayed in Eq. \eq{A41A42}. The last, crucial term
$A_{4,3}(\gamma)$, was related to the 4PM contribution in the local hyperbolic radial action via Eq. \eq{A43vsIr3}.
Using the latter relation we find
\be \label{A43bis}
A_{4,3}(\gamma) = A_{4,3}^{\pi^2}(\gamma) + A_{4,3}^{\rm rem}(\gamma)\,,
\ee
where the exact expression of the $\pi^2$ piece follows from the results of Refs. \cite{Bern:2021yeh,Dlapa:2024cje},
namely, 
\beq \label{A43pi2}
A_{4,3}^{\pi^2}(\gamma)=\frac12 {\mathcal M}_4^{\pi^2}\,,
\eeq
where the value of ${\mathcal M}_4^{\pi^2}$ was recalled in Eqs. \eq{M4pi2}, \eq{coefM4pi2}.
As the last piece in $A_{4,3}(\gamma)$ is obtained here as a limited PN series, let us,
for practical EOB purposes, display its PN expansion up to $O(\pinf^{30})$:
\bea
A_{4,3}^{\pi^2}(\g)&=&   \pi^2 \left[  -  \frac{41}{32}  
 -  \frac{33601}{6144}  p_\infty^2 \right.\nonumber\\
&-&  \frac{93031}{12288} p_\infty^4  -  \frac{9733841}{4194304} p_\infty^6\nonumber\\ 
&-&  \frac{303760055}{201326592}  p_\infty^8 + \frac{15380987189}{12884901888}  p_\infty^{10}\nonumber\\ 
&-& \frac{16541063287}{17179869184}  p_\infty^{12} 
+\frac{27948129277267}{35184372088832} p_\infty^{14} \nonumber\\
&-& \frac{752705766645767}{1125899906842624}  p_\infty^{16} \nonumber\\
&+& \frac{123692590730776925}{216172782113783808} p_\infty^{18} \nonumber\\
&-& \frac{107367805309183991}{216172782113783808} p_\infty^{20} \nonumber\\
&+& \frac{193147770737983523995}{442721857769029238784} p_\infty^{22} \nonumber\\ 
&-& \frac{2742100258254883550063}{7083549724304467820544} p_\infty^{24} \nonumber\\
&+&  \frac{157075421780003581656641}{453347182355485940514816 } p_\infty^{26}  \nonumber\\
&-& \frac{566610437031857627233835}{1813388729421943762059264} p_\infty^{28}   \nonumber\\
&+& \left. \frac{1404532330880150975098811995}{4951760157141521099596496896} p_\infty^{30}  \right]\nonumber\\
&+& O\left(p_\infty^{32}\right)\,,
\eea
The  remaining (rational) contribution $A_{4,3}^{\rm rem}(\gamma)$ is obtained by inserting our new
results in  Eq. \eq{A43vsIr3}. We find, to order $\pinf^{30}$,
\bea \label{A43rem}
A_{4,3}^{\rm rem}(\g)
&=&   \frac{10}{p_\infty^2} 
+\frac{767}{6}   + \frac{4033}{18}   p_\infty^2 \nonumber\\
&+&  \frac{6514457}{50400}  p_\infty^4  - \frac{6859063}{235200}  p_\infty^6\nonumber\\ 
&+&\frac{2233281361}{55883520}   p_\infty^8   -\frac{7951126982609}{319653734400}   p_\infty^{10} \nonumber\\
&+&  \frac{88416122577}{6412806400}  p_\infty^{12}  -\frac{1793449303541}{419404677120} p_\infty^{14} \nonumber\\
&-&\frac{5454709249034730007}{1168273537160970240}  p_\infty^{16} \nonumber\\
&+&  \frac{100602741876231302573}{7399065735352811520}p_\infty^{18} \nonumber\\
&-&  \frac{88377833003445193499}{3867693452570787840 }  p_\infty^{20}\nonumber\\
&+&  \frac{10228674629587944921234863}{313128461920130983526400} p_\infty^{22} \nonumber\\
&-& \frac{1353288120259581811168531699}{31312846192013098352640000} p_\infty^{24}\nonumber\\
&+&  \frac{1259141982075541949023240559}{23043194143182571438080000} p_\infty^{26}  \nonumber\\
&-&\frac{2364613958836747116924289119153131}{35266631604709254644952465408000}  p_\infty^{28} \nonumber\\  
&+&  \frac{880430137242960727600628224263303433}{10932655797459868939935264276480000} p_\infty^{30} \nonumber\\
&+&O\left(p_\infty^{32}\right)\,.
\eea
The first few terms in the $\pinf^2$ expansions of $A_{4,3}^{\pi^2}(\gamma)$ and $A_{4,3}^{\rm rem}(\gamma)$
agree with our previous 6PN-accurate results, Eqs. \eq{A43rem6PN}, \eq{A43pi26PN}. Let us also note that
the above results are directly applicable to the dynamics of bound states, using $\pinf^2 = \g^2-1$ with
the effective EOB energy $\Ef= - P_0$ being related to the real cm energy via Eqs. \eq{EvsEf}, \eq{Efvsg}.
Let us also mention that the apparent  $\g  \to 1$ singularity,  $ \frac{10}{p_\infty^2} $,  featuring in 
$A_{4,3}^{\rm rem}(\g)$ is spurious because it is compensated by the  $A_{4,1}(\gamma)$, and $A_{4,2}(\gamma)$
contributions. The full function $q_4^{\rm loc}(\g,\nu)$, Eq, \eq{q4loc}, is smooth (and analytic) at $\g=1$.

\section{Outlook}

The purely local-in-time $O(G^4)$ contribution to the conservative dynamics of binary systems has been 
obtained here. We encapsulated the gauge-invariant content of the 4PM dynamics in two different functions:
(i) the local part of the radial action, applicable (modulo a factor 2) both to hyperboliclike and to ellipticlike motions;
see Sec. \ref{4PMradialaction}; and (ii) the coefficient $q^{\rm loc}_4(\g,\nu)$  of the $\left(\frac{GM}R\right)^4$
contribution to the Effective One-Body Hamiltonian in energy gauge. See Eq. \eq{q4loc}, where the three
energy-dependent coefficients $A_{4,1}(\gamma)$, $A_{4,2}(\gamma)$, $A_{4,3}(\gamma)$ are
given by Eqs. \eq{A41A42}, \eq{A43}, \eq{A43pi2} and by Eq. \eq{A43rem}.

It would be pleasing to complete our results by computing the exact value of the $W_2^{\nu^2}(\g)$ integral,
Eq. \eq{W2final}, which we  here computed to $O(\pinf^{30})$. The latter integral contains only one scale.
It would also be nice to compute the full version of the (two-scale) $W_2(\g,\nu)$ integral, Eq. \eq{W2Aframe}.
Let us note  a few facts concerning the extra information that such exact PM results would bring.

Concerning the exact computation of $W_2(\g,\nu)$ (taken together with the corresponding exact result
for $W_1(\g,\nu)$ \cite{Dlapa:2024cje}) this would allow one to compute the 4PM part of the flexibility factor
$f(t)$. Our past work has derived the 6PN accurate value of $f(t)$, i.e. the values of the coefficients entering
its PN expansion, namely (using scaled variables $p_i=P_i/\mu$, $\frac1{r}= \frac{GM}{R}$)
\bea
f(t)&=& 1 + \frac{\nu}{c^2} \left(c_1 p_r^2+ c_2 p^2 + c_3 \frac1{r}\right)\nonumber\\ 
&+&  \frac{\nu}{c^4} \left(d_1(\nu) p_r^4+ d_2(\nu) p^4 +d_3(\nu) \frac{1}{r^2}\right.\nonumber\\
&+&\left. d_4(\nu) p^4p_r^2+d_5(\nu)\frac{ p_r^2}{r}+d_6(\nu) \frac{p^2}{r}\right)\,, 
\eea
see Sec. VII of Ref. \cite{Bini:2020hmy}.

The knowledge of the full integral $W_2(\g,\nu)$ would give the complete sequence of the $G$-independent part
of $f(t)$, namely
\bea
f(t)&=& 1 + \frac{\nu}{c^2} (c_1 p_r^2+ c_2 p^2 ) \nonumber\\
&+&  \frac{\nu}{c^4} (d_1(\nu) p_r^4+ d_2(\nu) p^4 + d_4(\nu)  p^2 p_r^2) \\ \nonumber
&+& \nu \sum_{k\geq 3} O(p^{2 k})\,.
\eea
However, such an information would still lack the knowledge of all the $G$-dependent terms 
(parametrized by $c_3, d_3, d_5, d_6$) which enter the bound-state local dynamics on par with the purely
velocity-dependent ones. 

The determination of the exact value of the simpler one-scale integral  $W_2^{\nu^2}(\g)$,
Eq. \eq{W2final}, would bring a valuable information. The corresponding structure of the
``remaining" (non $\pi^2$) contribution to the total radial action (given by the last term in Eq. (3)
of \cite{Bern:2021yeh}) is a combination, with rational coefficients, of
transcendental functions of weight up to 2, such as $\ln^2 (\frac{\g+1}{2})$
or ${\rm Li}_2(\frac{1-\g}{1+\g} )$. We  expect $W_2^{\nu^2}(\g)$ (and the
corresponding contributions to the local dynamics)  to have a similar
structure, and, in particular, to have, as closest singularity to the slow-motion limit $\g=1$,
a singularity at $\g=0$. In terms of the expansion in powers of $\pinf^2 = \g^2-1$, or, equivalently,
in powers of
\be
x \equiv 1- \g^2 = - \pinf^2
\ee
this would mean a singularity at $x= +1$, and a corresponding convergence radius equal to 1.

The high-order velocity expansions we derived above for the remaining parts of the local dynamics seem to
confirm this expectation. In particular, we considered the $x$-expansion of  $A_{4,3}^{\rm rem}(\g)$,
rescaled by its first term $\frac{10}{\pinf^2}= - \frac{10}{x}$, i.e
\bea \label{xexp}
{\widehat A}_{4,3}^{\rm rem}(x) &\equiv& \frac{\pinf^2}{10}A_{4,3}^{\rm rem}(\g)\nonumber\\
&=& 1 - \frac{767}{60} x +\frac{4033}{80} x^2 +\cdots \\ \nonumber
&=& \sum_{n =0}^{16} c_n x^n\,.
\eea
We found that the coefficients $c_n$ in this expansion seem to behave (after $n \gtrsim 5$)
like $c_n \approx a n + b $ with $ a \approx \frac{10}{9}$ and $b \approx - 10$. 
Such a mild growth indeed suggests a singularity at $x=+1$ of the type $\sim \frac{a}{(1-x)^2} = \frac{a}{\g^4}$.
The presence of  such a singularity indicates that it might be useful to trade the  $x$-expansion of the function ${\widehat A}_{4,3}^{\rm rem}(x)$
for the $x$-expansion of the function ${\check A}_{4,3}^{\rm rem}(x) \equiv (1-x)^2 {\widehat A}_{4,3}^{\rm rem}(x) $, i.e. to write $A_{4,3}(x)$
(with $x=1-\g^2$) as~\footnote{There might be better choices of the power of $(1-x)$ for improving the convergence.}
\beq
A_{4,3}(x)=- \frac{10}{x (1-x)^2} {\check A}_{4,3}^{\rm rem}(x)\,.
\eeq
One indeed finds that the coefficients of the $x$-expansion of ${\check A}_{4,3}^{\rm rem}(x)$ are  $O(50)$ for the first five terms and then
become  $O(0.1)$.

From the practical point of view, and notably for the use of our results to define an EOB Hamiltonian, the mild growth
of the coefficients of the $x$ expansion of the remaining contribution to the  $q^{\rm loc}_4(\g,\nu)$ function
suggests that our present, $O(\pinf^{30})$, results are sufficiently accurate to satisfactorily describe the local
4PM dynamics for the bound state motions of most relevance to the modelling of gravitational-wave templates.
Indeed, in the EOB approach one uses the two-body Hamiltonian description up to the crossing of some effective
lightring (corresponding to the peak of the waveform). In state-of-the-art EOB codes, this corresponds\footnote{We thank Alessandro Nagar for information on the minimum value of $\g$.} to a minimum value, $\g_{\rm min}$, of $\g$ , of order
$\g_{\rm min} \approx 0.82$. Such a value corresponds to a maximum value $x_{\rm max} = 1- \g_{\rm min}^2 \approx
0.33$. This value is sufficiently far from the expected singularity at $x=1$ for one to expect that our 
 $O(\pinf^{30})$ results will define a sufficiently accurate Hamiltonian dynamics. For instance,
 the last term in the expansion \eq{xexp}  of ${\widehat A}_{4,3}^{\rm rem}(x)$ is $\approx 8\times  (0.33)^{16} \approx 1.6 \times 10^{- 7}$.
And the last term in the expansion of $ {\check A}_{4,3}^{\rm rem}(x)$ is $\approx 0.1 \times  (0.33)^{16} \approx 2 \times 10^{- 9}$.
In addition, we have checked the numerical accuracy of our velocity-expanded results
by numerically computing the integral  $W_2^{\nu^2}(2b)$, Eq. \eq{W2final}, when injecting
purely imaginary values of $v= i v_I$ in the algebraic expression of the integrand. The
corresponding integration on (real) values of $\theta_A, \phi_A$ and $T_A$ is well defined
up to $v_I = 1$, which corresponds to $x=1-\g^2=\frac12$.

Let us finally mention another (practical)  limitation of our result on the 4PM local dynamics. As recalled in
Eq. \eq{Stot}, when applying the local Hamiltonian $H^{G^4}_{\rm loc}$ to the description of
bound systems (of most immediate interest for gravitational-wave purposes) it must be completed
both by the $O(G^5)$,  $O(G^6)$, and  $O(G^7)$ {\it local Hamiltonian} terms contained in our previous 6PN
local Hamiltonian and by bound-state version of the 6PN-accurate  {\it nonlocal Hamiltonian}  \cite{Bini:2020hmy}.
In view of the nonperturbative-in-$G$ nature of the bound-state nonlocal dynamics, it is unclear how this
dynamics could, eventually, be upgraded to some full PM accuracy. One can, however, hope, that the inclusion
in the local dynamics of the very-high-PN-order terms obtained above might improve the descrition of bound-state
dynamics (see \cite{Buonanno:2024byg} for recent work in this direction).

\appendix 

\section{Changing frames in two-body scattering} 
\label{AppFrames}

In the (leading-order) Post-Minkowskian description of a two body system, say A (with mass $m_A \equiv m_1$) 
and B (with mass $m_B \equiv m_2$) , several frames appear as natural, notably, the rest frame of body A, that of body B, and the center of mass (cm) frame (we will not use the center of velocity (cv) frame). At the leading PM order the
bodies move on straight worldlines.
 We  use the mostly plus signature.
 
 The world lines of the two bodies, parametrized by proper times $\tau_A$ and $\tau_B$, read
\bea
\label{worldlines}
z^\mu_A(\tau_A)&=&z^\mu_A(0)+u^\mu_A\tau_A\,,\nonumber\\
z^\mu_B(\tau_B)&=& z^\mu_B(0)+u^\mu_B\tau_B\,,
\eea
where $\tau_A=0$ and $\tau_B=0$ describe the points on the worldlines
such that $z^\mu_A(0)- z^\mu_B(0) = b  e^\mu_2$ (with $e^\mu_2$ a unit spatial vector) is orthogonal
to both worldlines. Here, $b >0$ is the impact parameter. In a general frame we have
\beq
z^\mu_A(0)=b_A e^\mu_2\,,\qquad z^\mu_B(0)=b_B e^\mu_2\,,
\eeq
with 
\beq
b_A-b_B=b\,,
\eeq
In the frame of body $A$ (or $A$-frame) we have $b_A=0$. Similarly, $b_B=0$ in the $B$-frame.
In the cm frame we have 
\bea
{\rm cm \; frame:}&&   E_A b_A+E_Bb_B=0\,, \nonumber\\ 
&\Rightarrow&  b_A=\frac{E_B}{E}b\,,\quad b_B=-\frac{E_A}{E}b\,.
\eea

We denote the Cartesian coordinates of the  rest frame of body A as $(t_A, x_A, y_A, z_A)$, with
associated polar coordinates $(t_A, r_A, \theta_A, \phi_A)$, and associated basis vectors $(e^\mu_0,e^\mu_1,e^\mu_2,e^\mu_3)$.
Within this frame, body A is at rest, 
\beq u^\mu_A=e^\mu_0,\eeq 
while the orbit of body B lies in the $e^\mu_1-e^\mu_2$ plane with initial velocity in
the $e_1$-direction with velocity $v$,
\beq
u^\mu_B=\gamma (e^\mu_0+v e^\mu_1)\,.
\eeq
The unit spatial vector pointing from the origin,  $(t_A, x_A, y_A, z_A)=(0,0,0,0)$, toward the observer is
\bea
{\mathbf n}&=&\cos\theta_A e_1+\sin \theta_A \cos\phi_A e_2 +\sin \theta_A \sin\phi_A e_3\nonumber\\
&\equiv & \alpha_A e_1+\beta_A e_2 +\delta_A e_3\,.
\eea

The coordinates of the other frames are respectively denoted
 $(t_B, x_B, y_B, z_B)$ and  $(t_{\rm cm}, x_{\rm cm}, y_{\rm cm}, z_{\rm cm})$.

Let us recall some useful relations.
\beq
p_\infty= \sqrt{\g^2-1}=\gamma v\,,
\eeq
as well as the relation between the (rescaled) angular momentum $j=J/(G m_A m_B)$ and the impact parameter $b$
\beq
\frac{GM}{b}=\frac{p_\infty}{hj}\,,\qquad
\frac{1}{j}=\frac{GMh}{p_\infty b}\,,
\eeq
and the expression of the total cm energy 
\beq
E\equiv E_{\rm cm}=\sqrt{m_A^2+m_B^2+2\gamma m_Am_B}\equiv Mh\,,
\eeq
with $M=m_A+m_B$.

The Lorentz transformation (with velocity $v$) from the A frame to the B one corresponds to
\bea
t_B&=&\gamma (t_A -vx_A)\,,\qquad x_B=\gamma (x_A-vt_A)\,,\nonumber\\
y_B&=&y_A-b\,,\qquad z_B=z_A\,,
\eea
and implies ($r_X$ denoting the distance to the origin in the $X$-frame)
\beq
r_B=\gamma (1-\alpha_A v)r_A\,,\qquad u_{{\rm ret}}^B=\frac{u_{{\rm ret}}^A+\beta_A b}{\gamma (1-\alpha_A v)}\,,
\eeq
where we introduced the two retarded times $u_{{\rm ret}}^A=t_A-r_A$, $u_{{\rm ret}}^B=t_B-r_B$,   assuming
 $b\ll r_A$, $|u_{{\rm ret}}^A|\ll r_A$, $|u_{{\rm ret}}^B|\ll r_B$.
Similarly, passing from the frame A to the cm one corresponds to the boost velocity $v_{\rm cm}$ such that
\bea
U^\mu\equiv u^\mu_{\rm cm}&=&\frac{m_A u^\mu_A +m_B u^\mu_B}{E}\nonumber\\
&=& \frac{(m_A  +m_B\gamma)e^\mu_0+ m_B \gamma v e^\mu_1}{E}\nonumber\\
&=& \gamma_{\rm cm} (e^\mu_0+v_{\rm cm}e^\mu_1)\,,
\eea
where
\beq
v_{\rm cm}=\frac{m_B \gamma v}{m_A  +m_B\gamma}\,,\qquad \gamma_{\rm cm}=\frac{m_A  +m_B\gamma}{E}\,.
\eeq
Let us recall the useful relations
\beq
\frac{E_A}{m_A}=\frac{m_A+\gamma m_B}{E}\,,\qquad \frac{E_B}{m_B}=\frac{m_B+\gamma m_A}{E}\,,
\eeq
which imply 
\beq
\gamma_{\rm cm}=\frac{E_A}{m_A}\,.
\eeq
The latter relation shows that $v_{\rm cm}$ and $\gamma_{\rm cm}$ are dissymetric functions
of the two masses (though this is not indicated by the notation).

One can easily find the relations among $\tau_A$, $\tau_B$ and $t_{\rm cm}$, when slicing the two worldlines
by the $t_{\rm cm}= {\rm cst}$ hyperplanes. In fact, using Eqs. \eqref{worldlines}, the condition
\beq
U\cdot  z_A(\tau_A)=-t_{\rm cm}\,,
\eeq
implies the following relation between $\tau_A$  and $t_{\rm cm}$
\beq
t_{\rm cm}=\frac{E_A}{m_A}\tau_A\,,
\eeq
and, similarly,
\beq
t_{\rm cm}=\frac{E_B}{m_B}\tau_B\,.
\eeq

The Lorentz transformation  from the A-frame to the cm one reads
\bea
t_{\rm cm}&=&\gamma_{\rm cm} (t_A -v_{\rm cm}x_A)\,,\qquad x_{\rm cm}=\gamma_{\rm cm} (x_A-v_{\rm cm}t_A)\,,\nonumber\\
y_{\rm cm}&=&y_A-  b_A \,,\qquad z_{\rm cm}=z_A\,,
\eea
and implies
\bea
\label{boost_A_cm}
r_{\rm cm}&=&\gamma_{\rm cm} (1-\alpha_A v_{\rm cm})r_A\,,\nonumber\\ 
u_{{\rm ret}}^{\rm cm}&=& \frac{u_{{\rm ret}}^A+\beta_A b_A}{\gamma_{\rm cm} (1-\alpha_A v_{\rm cm})}\,,
\eea
with a similar one corresponding to the boost from B to cm, where instead of $b_A$ one  has $b_B$.

Eqs. \eqref{boost_A_cm}  shows how the link between $u_{{\rm ret}}^A$ and $u_{{\rm ret}}^{\rm cm}$  depends on the angles and implies,
\beq
\label{u_ret_cm}
u_{{\rm ret}}^A=\gamma_{\rm cm} (1-\alpha_A v_{\rm cm}) u_{{\rm ret}}^{\rm cm} -\beta_A b_A\,.
\eeq
We will use the notation
\bea
f_{\rm cm}&=&\gamma_{\rm cm} (1-\alpha_A v_{\rm cm} )
=\frac{1}{\gamma_{\rm cm} (1+\alpha_{\rm cm} v_{\rm cm} )}\,, \qquad
\eea 
so that Eq. \eqref{u_ret_cm} becomes
\beq
u_{{\rm ret}}^A=f_{\rm cm} \, u_{{\rm ret}}^{\rm cm} -\beta_A b_A\,.
\eeq

The transformation of the polar angles is  given by
\beq
\label{alpha_A_vs_cm}
\alpha_A=\frac{\alpha_{\rm cm}+v_{\rm cm}}{1+\alpha_{\rm cm}v_{\rm cm}}\,,\qquad \phi_A=\phi_{\rm cm}\,,
\eeq
with
\bea
\alpha_{\rm cm}&=& \frac{\alpha_A-v_{\rm cm}}{1-\alpha_A v_{\rm cm}}\,, \nonumber\\
\beta_{\rm cm}&=& \frac{\beta_A}{\gamma_{\rm cm}(1-\alpha_A v_{\rm cm})}\,, \nonumber\\
\delta_{\rm cm}&=& \frac{\delta_A}{\gamma_{\rm cm}(1-\alpha_A v_{\rm cm})} \,,
\eea
and 
\beq
\label{partial_A}
\left[\frac{\partial \alpha_A}{\partial \alpha_{\rm cm}},\frac{\partial \beta_A}{\partial \alpha_{\rm cm}},\frac{\partial \delta_A}{\partial \alpha_{\rm cm}} \right]
=[f_{\rm cm}^2,f_{\rm cm},f_{\rm cm}]\,.
\eeq
Finally, the volume element transforms as
\beq
\label{4_volumes}
|du_{{\rm ret}}^A\wedge d^2\Omega_A|=f_{\rm cm}^3\, |d u_{{\rm ret}}^{\rm cm}\wedge d^2\Omega_{\rm cm}| \,,
\eeq
since by using Eq. \eqref{partial_A} one gets
\bea
d^2\Omega_A&=&\sqrt{1-\alpha_A^2}d\theta_A\wedge d\phi_A\nonumber\\
&=& -d\alpha_A \wedge d\phi_A\nonumber\\
&=& -\frac{\partial \alpha_A}{\partial \alpha_{\rm cm}}d\alpha_{\rm cm}\wedge d\phi_{\rm cm}\nonumber\\
&=&  + f_{\rm cm}^2 d^2\Omega_{\rm cm}
\eea
and from Eq. \eqref{u_ret_cm}
\beq
du_{{\rm ret}}^A =f_{\rm cm} du_{{\rm ret}}^{\rm cm}-\gamma_{\rm cm}v_{\rm cm}u_{\rm cm}d\alpha_A-b_Ad\beta_A\,, 
\eeq
so that
\bea
du_{{\rm ret}}^A \wedge d^2\Omega_A&=&f_{\rm cm} du_{{\rm ret}}^{\rm cm}\wedge d^2\Omega_A\,.
\eea

It is convenient to introduce new, rescaled retarded-time variables:
\beq
T_A  
=\frac{\gamma v}{b}u_{{\rm ret}}^A\,,\qquad
T_B  
=\frac{\gamma v}{b}u_{{\rm ret}}^B \,.
\eeq
The relation between the retarded times, Eq. \eqref{boost_A_cm}, becomes then
\beq
\label{TB_vs_TA}
T_B=\frac{T_A+\gamma v \beta_A}{\gamma (1-v \alpha_A)}\,, 
\eeq
with inverse
\beq
\frac{T_A}{\gamma}=T_B \left(1-v\alpha_A\right)-v \beta_A\,.
\eeq
We also define  
\beq
T_{\rm cm}=\frac{\gamma v}{b}\frac{m_A m_B}{E_A E_B} t_{\rm cm}\,,
\eeq
as well as the distance between the two bodies in the cm frame at a fixed $t_{\rm cm}$, $r_{\rm cm}=|z_A(\tau_A(t_{\rm cm}))-z_B(\tau_B(t_{\rm cm}))|$
\bea
r_{\rm cm}^2&=&b^2+\left( \frac{m_Am_B}{E_AE_B}\right)^2\gamma^2v^2 t_{\rm cm}^2\nonumber\\
&=& b^2(1+T_{\rm cm}^2)\,.
\eea

\section{Determination of $W_2$}
\label{W2_details}

We have obtained our  result for $W_2^{\nu^2}$ via two
independent computations, one based on  Ref. \cite{kov-tho4}, and one based on Ref. \cite{Jakobsen:2021smu}.
We have also partially checked our results by using the spurious-pole-free waveform derived in Ref. \cite{DeAngelis:2023lvf}.
Let us indicate a few relevant details of the approaches we used.

\subsection{Determination of $W_2$ following Ref. \cite{kov-tho4}}

The initial configuration of the scattering process is characterized by the parameter $v$ representing the relative velocity of the two bodies in the rest frame of body $A$. Therefore, the amplitudes of the emitted waves for both polarizations $+,\times$ at a given spacetime point depend on the  coordinates $(\theta_A, \phi_A)$ of this point  and the retarded time $u_A=t_A-r_A$,
\beq
\lim_{r_A\to \infty}( r_A  h^A_{+,\times})=A_{+,\times}(u_A,\theta_A,\phi_A; v)\,,
\eeq
and their expressions are given Ref. \cite{kov-tho4}. Passing to $T_A= \frac{v\gamma}{b}u_A$
The radiated gravitational wave energy flux is then 
\bea
F_{E}^A(T_A,\theta_A,\phi_A; v)&=&\frac{1}{16\pi} \frac{\gamma v}{b}\frac{G^2 m_1^2 m_2^2}{b^2}\times \nonumber\\
&& \left[ \left(\frac{d\widehat A_+}{dT_A}\right)^2+\left(\frac{d\widehat  A_\times}{dT_A}\right)^2\right]\,,\qquad
\eea
where the $\frac{\gamma v}{b}$  converts from $u_A$ to $T_A$ and a  coefficient $(Gm_1m_2)/b$ has been has been factored out of both $A_+$ and $A_\times$,
\beq
A_{+,\times}=\frac{Gm_1m_2}{b} \widehat A_{+,\times}\,. 
\eeq
The explicit expression of   $F_{E}^A$ leads to a large expression.
However, expanding $F_{E}^A$ in powers of $v$ (up to a high power) allows one to handle  this expression,
and then to  integrate over the angles to get the instantaneous power: 
\beq
F_{E}^{A {\rm int}}(T_A; v)=\int \sin\theta_A d\theta_A d\phi_A \, F_{E}^A(T_A,\theta_A,\phi_A; v)\,.
\eeq
The beginning of the velocity expansion of $F_{E}^{A {\rm int}}(T_A; v)$ reads
\bea
F_{E}^{A {\rm int}}(T_A; v)&=& \sum_{k\geq0} F_{EA,k}(T_A) v^{2k+1} \nonumber\\
&=&\frac{G^2 m_1^2 m_2^2}{b^3}\left[ \frac{8}{15}\frac{(T_A^2 + 12)}{(T_A^2 + 1)^3}v\right.\nonumber\\
&-&\frac{2}{105} \frac{86 T_A^4+606 T_A^2-617}{  (T_A^2 + 1)^4 }\, v^3\nonumber\\
&+&\left. O(v^5)\right]\,.
\eea
As indicated,  the even powers of $v$ disappear after integrating over angles.
The radiated energy ${\mathcal E}$ follows by integrating this expression over $T_A$ 
\bea
\int_{-\infty}^\infty dT_A F_{E}^{A {\rm int}}(T_A; v)&=&\frac{G^2 m_1^2 m_2^2}{b^3} \left[\frac{37}{15}\pi v \right. \nonumber\\
&+&\left.\frac{2393}{840}\pi v^3+O(v^5)\right]\,.
\eea
The   part of the action $W_2$ we are interested in computing, instead, is obtained by multiplying $F_{E}^{A {\rm int}}(T_A; v)$ by the logarithm of the relative distance of the two bodies modulo a scale factor which we can conveniently
choose as $s=2b$,
\beq
\ln\left(\frac{1}{4}\left(1+\frac{T_A^2}{\gamma^2}\right)\right)\,.
\eeq
The latter is then expanded in $v$ with the further replacement 
\beq
\ln(1+T_A^2)=\frac{(1+T_A^2)^\epsilon-1}{\epsilon}\,,
\eeq
so that the limit $\epsilon \to 0$ will give back the original logarithm. We find for
\beq
{\mathcal W}_2(T_A; v)=F_{E}^{A {\rm int}}(T_A; v) \ln\left(\frac{1}{4}\left(1+\frac{T_A^2}{\gamma^2}\right)\right)\bigg|_{v\, {\rm exp}}
\eeq
a velocity expansion starting as
\bea
{\mathcal W}_2(T_A; v)&=&
\frac{G^2 m_1^2 m_2^2}{b^3}\left\{ 
\frac{8}{15}\frac{(T_A^2 + 12)}{(T_A^2 + 1)^3}v\right.\nonumber\\
&-&\left.\frac{2}{105} \frac{86 T_A^4+606 T_A^2-617}{  (T_A^2 + 1)^4 }\, v^3 +O(v^5)
\right\}\times  \nonumber\\
&& \left[ \frac{(T_A^2 + 1)^\epsilon - 1}{\epsilon} -\ln(4)  -  \frac{T_A^2 }{(T_A^2 + 1)}v^2\right.\nonumber\\
&+& \left. O(v^4)\right]\,,
\eea
which after integration and after taking the limit $\epsilon \to 0$ gives $W_{2}^{s=2b}$, Eq. \eqref{W20sf}.


\subsection{Determination of $W_2$ following Ref. \cite{Jakobsen:2021smu}}

Ref. \cite{Jakobsen:2021smu} works also in the  $A$-frame but uses as time variable the quantity $\tau$
related to the variables of the previous subsection as
\beq
\tau = v \g u^B_{\rm ret} = b \, T_B \,,
\eeq
see Eq. (18) there. 
A special characteristic of the waveform expression of \cite{Jakobsen:2021smu} is that they concentrate the
delicate spurious pole $\frac1{S^2}$ present all over the expressions of \cite{kov-tho4} into a particular
integral over an auxiliary  Feynman  parameter $\alpha$  of the form
\beq
4\pi I_{ij}=\int_0^1d\alpha \frac{{\mathcal I}_{ij}(\alpha)}{G(\alpha)^{3/2}}\,,
\eeq
where the denominator is quadratic in $\alpha$,
\beq
G(\alpha)=G_0+2\alpha VG_1+\alpha^2 V^2 G_2\,.
\eeq
Here $V=\frac{v}{1-v\cos\theta_A}$ and (setting $b=1$ hereafter)
\bea
G_0&=& 1+\tau^2\,,\nonumber\\
G_1&=&\cos\theta_A -\sin \theta_A \cos \phi_A \tau\,, \nonumber\\
G_2&=&-\sin^2 \theta_A \sin^2\phi_A   \,.
\eea

The numerator has the structure:
\beq
{\mathcal I}_{ij}(\alpha)={\rm det}(M)[M^{-1}_{bb} M^{-1}_{ij}-M^{-1}_{ib}M^{-1}_{jb}]\,,
\eeq
with $M_{ij}=\delta_{ij}+\alpha L_{ij}$ and we used the compact notation $A_{ij}b^j=A_{ib}$, $A_{ij}b^ib^j=A_{bb}$ for a generic matrix $A$.

The dependence of the numerator ${\mathcal I}_{ij}(\alpha)$ on $\alpha$ is linear:
\beq
{\mathcal I}_{ij}(\alpha)={\mathcal I}^{(0)}_{ij}(\tau) +\alpha {\mathcal I}^{(1)}_{ij}(\tau,\theta, \phi )\,,
\eeq
with ${\mathcal I}^{(0)}_{ij}$  quadratic in $\tau$ and ${\mathcal I}^{(1)}_{ij}$ linear in $\tau$,
and depending on the polar angles $\theta_A, \phi_A $.

Therefore the integral $4\pi I_{ij}$ is reduced to the following two $\alpha$ integrals
\bea
{\mathcal G}_0&=&\int_0^1   \frac{d\alpha}{G(\alpha)^{3/2}}\,,\nonumber\\
{\mathcal G}_1&=&\int_0^1   \frac{\alpha d\alpha}{G(\alpha)^{3/2}}\,.
\eea
These integrals are elementary but effecting them introduces the spurious pole $S^2$ in the
guise of the discriminant $\Delta(G)=4(G_1^2 - G_0 G_2)$ of the quadratic polynomial $G(\alpha)$.

One can,however, avoid generating the latter spurious pole by expanding in powers of $V$ the integrand
$G(\alpha)^{-3/2}$ before integrating over $\alpha$. The latter expansion generates a fast-evaluated series
of Gegenbauer polynomials.

We have used the resulting expansions to have an independent check of the results obtained directly from
the original Kovacs-Thorne form of the waveform discussed in the previous subsection.

\section*{Acknowledgements}
The present research was partially supported by the 2021
Balzan Prize for Gravitation: Physical and Astrophysical Aspects, awarded to T. Damour.
We thank Stefano de Angelis, Andrea Geralico and Alessandro Nagar for informative discussions.
D.B.  
acknowledges sponsorship of the Italian Gruppo Nazionale per la Fisica Matematica (GNFM)
of the Istituto Nazionale di Alta Matematica (INDAM), 
as well as the hospitality and the highly stimulating environment of the Institut des Hautes Etudes Scientifiques.


\begin{thebibliography}{99}

\bibitem{Bini:2019nra}
D.~Bini, T.~Damour and A.~Geralico,
``Novel approach to binary dynamics: application to the fifth post-Newtonian level,''
Phys. Rev. Lett. \textbf{123}, no.23, 231104 (2019)
[arXiv:1909.02375 [gr-qc]].

\bibitem{Bini:2020wpo}
D.~Bini, T.~Damour and A.~Geralico,
``Binary dynamics at the fifth and fifth-and-a-half post-Newtonian orders,''
Phys. Rev. D \textbf{102}, no.2, 024062 (2020)
[arXiv:2003.11891 [gr-qc]].

\bibitem{Bini:2020nsb}
D.~Bini, T.~Damour and A.~Geralico,
``Sixth post-Newtonian local-in-time dynamics of binary systems,''
Phys. Rev. D \textbf{102}, no.2, 024061 (2020)
[arXiv:2004.05407 [gr-qc]].

\bibitem{Damour:2019lcq}
T.~Damour,
``Classical and quantum scattering in post-Minkowskian gravity,''
Phys. Rev. D \textbf{102}, no.2, 024060 (2020)
[arXiv:1912.02139 [gr-qc]].

\bibitem{Bern:2021dqo}
Z.~Bern, J.~Parra-Martinez, R.~Roiban, M.~S.~Ruf, C.~H.~Shen, M.~P.~Solon and M.~Zeng,
``Scattering Amplitudes and Conservative Binary Dynamics at ${\cal O}(G^4)$,''
Phys. Rev. Lett. \textbf{126}, no.17, 171601 (2021)
[arXiv:2101.07254 [hep-th]].

\bibitem{Damgaard:2023ttc}
P.~H.~Damgaard, E.~R.~Hansen, L.~Plant\'e and P.~Vanhove,
``Classical observables from the exponential representation of the gravitational S-matrix,''
JHEP \textbf{09}, 183 (2023)
[arXiv:2307.04746 [hep-th]].

\bibitem{Bern:2024adl}
Z.~Bern, E.~Herrmann, R.~Roiban, M.~S.~Ruf, A.~V.~Smirnov, V.~A.~Smirnov and M.~Zeng,
``Amplitudes, Supersymmetric Black Hole Scattering at $\mathcal{O}(G^5)$, and Loop Integration,''
[arXiv:2406.01554 [hep-th]].

\bibitem{Bern:2021yeh}
Z.~Bern, J.~Parra-Martinez, R.~Roiban, M.~S.~Ruf, C.~H.~Shen, M.~P.~Solon and M.~Zeng,
``Scattering Amplitudes, the Tail Effect, and Conservative Binary Dynamics at O(G4),''
Phys. Rev. Lett. \textbf{128}, no.16, 161103 (2022)
[arXiv:2112.10750 [hep-th]].

\bibitem{Driesse:2024xad}
M.~Driesse, G.~U.~Jakobsen, G.~Mogull, J.~Plefka, B.~Sauer and J.~Usovitsch,
``Conservative Black Hole Scattering at Fifth Post-Minkowskian and First Self-Force Order,''
[arXiv:2403.07781 [hep-th]].

\bibitem{Dlapa:2024cje}
C.~Dlapa, G.~K\"alin, Z.~Liu and R.~A.~Porto,
``Local in Time Conservative Binary Dynamics at Fourth Post-Minkowskian Order,''
Phys. Rev. Lett. \textbf{132}, no.22, 221401 (2024)
[arXiv:2403.04853 [hep-th]].

\bibitem{Dlapa:2021vgp}
C.~Dlapa, G.~K\"alin, Z.~Liu and R.~A.~Porto,
``Conservative Dynamics of Binary Systems at Fourth Post-Minkowskian 
Order in the Large-Eccentricity Expansion,''
Phys. Rev. Lett. \textbf{128}, no.16, 161104 (2022)
[arXiv:2112.11296 [hep-th]].

\bibitem{Jakobsen:2023ndj}
G.~U.~Jakobsen, G.~Mogull, J.~Plefka, B.~Sauer and Y.~Xu,
``Conservative Scattering of Spinning Black Holes at Fourth Post-Minkowskian Order,''
Phys. Rev. Lett. \textbf{131}, no.15, 151401 (2023)
[arXiv:2306.01714 [hep-th]].

\bibitem{Damour:2016gwp}
T.~Damour,
``Gravitational scattering, post-Minkowskian approximation and Effective One-Body theory,''
Phys. Rev. D \textbf{94}, no.10, 104015 (2016)
[arXiv:1609.00354 [gr-qc]].

\bibitem{Damour:2016gwp}
T.~Damour,
``Gravitational scattering, post-Minkowskian approximation and Effective One-Body theory,''
Phys. Rev. D \textbf{94}, no.10, 104015 (2016)
[arXiv:1609.00354 [gr-qc]].

\bibitem{Damour:2017zjx}
T.~Damour,
``High-energy gravitational scattering and the general relativistic two-body problem,''
Phys. Rev. D \textbf{97}, no.4, 044038 (2018)
doi:10.1103/PhysRevD.97.044038
[arXiv:1710.10599 [gr-qc]].

\bibitem{Cheung:2018wkq}
C.~Cheung, I.~Z.~Rothstein and M.~P.~Solon,
``From Scattering Amplitudes to Classical Potentials in the Post-Minkowskian Expansion,''
Phys. Rev. Lett. \textbf{121}, no.25, 251101 (2018)
[arXiv:1808.02489 [hep-th]].

\bibitem{Kalin:2019rwq}
G.~K\"alin and R.~A.~Porto,
``From Boundary Data to Bound States,''
JHEP \textbf{01}, 072 (2020)
[arXiv:1910.03008 [hep-th]].

\bibitem{Antonelli:2019ytb}
A.~Antonelli, A.~Buonanno, J.~Steinhoff, M.~van de Meent and J.~Vines,
``Energetics of two-body Hamiltonians in post-Minkowskian gravity,''
Phys. Rev. D \textbf{99}, no.10, 104004 (2019)
[arXiv:1901.07102 [gr-qc]].

\bibitem{Cristofoli:2019neg}
A.~Cristofoli, N.~E.~J.~Bjerrum-Bohr, P.~H.~Damgaard and P.~Vanhove,
``Post-Minkowskian Hamiltonians in general relativity,''
Phys. Rev. D \textbf{100}, no.8, 084040 (2019)
[arXiv:1906.01579 [hep-th]].

\bibitem{Khalil:2022ylj}
M.~Khalil, A.~Buonanno, J.~Steinhoff and J.~Vines,
``Energetics and scattering of gravitational two-body systems at fourth post-Minkowskian order,''
Phys. Rev. D \textbf{106}, no.2, 024042 (2022)
[arXiv:2204.05047 [gr-qc]].

\bibitem{Bini:2018ywr}
D.~Bini and T.~Damour,
``Gravitational spin-orbit coupling in binary systems at the second post-Minkowskian approximation,''
Phys. Rev. D \textbf{98}, no.4, 044036 (2018)
[arXiv:1805.10809 [gr-qc]].

\bibitem{Damour:2022ybd}
T.~Damour and P.~Rettegno,
``Strong-field scattering of two black holes: Numerical relativity meets post-Minkowskian gravity,''
Phys. Rev. D \textbf{107}, no.6, 064051 (2023)
[arXiv:2211.01399 [gr-qc]].

\bibitem{Damour:2015isa}
T.~Damour, P.~Jaranowski and G.~Sch\"afer,
``Fourth post-Newtonian effective one-body dynamics,''
Phys. Rev. D \textbf{91}, no.8, 084024 (2015)
[arXiv:1502.07245 [gr-qc]].

\bibitem{Damour:2014jta}
T.~Damour, P.~Jaranowski and G.~Sch\"afer,
``Nonlocal-in-time action for the fourth post-Newtonian conservative dynamics of two-body systems,''
Phys. Rev. D \textbf{89}, no.6, 064058 (2014)
[arXiv:1401.4548 [gr-qc]].

\bibitem{Bini:2020hmy}
D.~Bini, T.~Damour and A.~Geralico,
``Sixth post-Newtonian nonlocal-in-time dynamics of binary systems,''
Phys. Rev. D \textbf{102}, no.8, 084047 (2020)
[arXiv:2007.11239 [gr-qc]].

\bibitem{Blanchet:1987wq}
L.~Blanchet and T.~Damour,
``Tail Transported Temporal Correlations in the Dynamics of a Gravitating System,''
Phys. Rev. D \textbf{37}, 1410 (1988)

\bibitem{Foffa:2011np}
S.~Foffa and R.~Sturani,
``Tail terms in gravitational radiation reaction via effective field theory,''
Phys. Rev. D \textbf{87}, no.4, 044056 (2013)
[arXiv:1111.5488 [gr-qc]].

\bibitem{Galley:2015kus}
C.~R.~Galley, A.~K.~Leibovich, R.~A.~Porto and A.~Ross,
``Tail effect in gravitational radiation reaction: Time nonlocality and renormalization group evolution,''
Phys. Rev. D \textbf{93}, 124010 (2016)
[arXiv:1511.07379 [gr-qc]].

\bibitem{Bernard:2017bvn}
L.~Bernard, L.~Blanchet, A.~Boh\'e, G.~Faye and S.~Marsat,
``Dimensional regularization of the IR divergences in the Fokker action of point-particle binaries at the fourth post-Newtonian order,''
Phys. Rev. D \textbf{96}, no.10, 104043 (2017)
[arXiv:1706.08480 [gr-qc]].

\bibitem{Foffa:2019yfl}
S.~Foffa, R.~A.~Porto, I.~Rothstein and R.~Sturani,
``Conservative dynamics of binary systems to fourth Post-Newtonian order in the EFT approach II: Renormalized Lagrangian,''
Phys. Rev. D \textbf{100}, no.2, 024048 (2019)
[arXiv:1903.05118 [gr-qc]].


\bibitem{Bern:2022jvn}
Z.~Bern, J.~Parra-Martinez, R.~Roiban, M.~S.~Ruf, C.~H.~Shen, M.~P.~Solon and M.~Zeng,
``Scattering amplitudes and conservative dynamics at the fourth post-Minkowskian order,''
PoS \textbf{LL2022}, 051 (2022)

\bibitem{kov-tho4}
S. J. Kovacs and K. S. Thorne, 
\lq\lq The Generation of Gravitational
Waves. 4. Bremsstrahlung," Astrophys. J. 224,
62 (1978) 


\bibitem{Jakobsen:2021smu}
G.~U.~Jakobsen, G.~Mogull, J.~Plefka and J.~Steinhoff,
``Classical Gravitational Bremsstrahlung from a Worldline Quantum Field Theory,''
Phys. Rev. Lett. \textbf{126}, no.20, 201103 (2021)
[arXiv:2101.12688 [gr-qc]].

\bibitem{Kosower:2018adc}
D.~A.~Kosower, B.~Maybee and D.~O'Connell,
``Amplitudes, Observables, and Classical Scattering,''
JHEP \textbf{02}, 137 (2019)
[arXiv:1811.10950 [hep-th]].

\bibitem{Mougiakakos:2021ckm}
S.~Mougiakakos, M.~M.~Riva and F.~Vernizzi,
``Gravitational Bremsstrahlung in the post-Minkowskian effective field theory,''
Phys. Rev. D \textbf{104}, no.2, 024041 (2021)
[arXiv:2102.08339 [gr-qc]].

\bibitem{DeAngelis:2023lvf}
S.~De Angelis, R.~Gonzo and P.~P.~Novichkov,
``Spinning waveforms from KMOC at leading order,''
[arXiv:2309.17429 [hep-th]].

\bibitem{Herrmann:2021lqe}
E.~Herrmann, J.~Parra-Martinez, M.~S.~Ruf and M.~Zeng,
``Gravitational Bremsstrahlung from Reverse Unitarity,''
Phys. Rev. Lett. \textbf{126}, no.20, 201602 (2021)
[arXiv:2101.07255 [hep-th]].

\bibitem{Herrmann:2021tct}
E.~Herrmann, J.~Parra-Martinez, M.~S.~Ruf and M.~Zeng,
``Radiative classical gravitational observables at $ \mathcal{O} $(G$^{3}$) from scattering amplitudes,''
JHEP \textbf{10}, 148 (2021)
[arXiv:2104.03957 [hep-th]].

\bibitem{Buonanno:2024byg}
A.~Buonanno, G.~Mogull, R.~Patil and L.~Pompili,
``Post-Minkowskian Theory Meets the Spinning Effective-One-Body Approach for Bound-Orbit Waveforms,''
[arXiv:2405.19181 [gr-qc]].


\end{thebibliography}
\end{document}